\PassOptionsToPackage{pdfpagelabels=false}{hyperref} 
\documentclass[fleqn,usenatbib]{mnras}
\usepackage{newtxtext,newtxmath}
\usepackage[T1]{fontenc}
\usepackage{ae,aecompl}

\usepackage{textcmds}
\usepackage{booktabs}
\usepackage{adjustbox}

\usepackage{graphicx}
\usepackage{braket}
\usepackage{amsmath}
\usepackage{times}

\newcommand{\kpc}                      {\,{\rm kpc}}
\newcommand{\Mpc}                      {\,{\rm Mpc}}

\newcommand{\pkpc}                      {\,{\rm{pkpc}}}
\newcommand{\pMpc}                      {\,{\rm pMpc}}

\newcommand{\ckpc}                      {\,{\rm ckpc}}
\newcommand{\cMpc}                      {\,{\rm cMpc}}

\newcommand{\Msun}                    {\,{\rm M}_\odot}

\newcommand{\Msunyrsqkpc}                  {\,{\rm M}_\odot\,{\rm yr}^{-1}\,{\rm kpc}^{-2}}

\defcitealias{hill21}{Paper~I}
 
\title[Intrinsic alignments of radio galaxies]
{Intrinsic alignments of the extended radio continuum emission of galaxies in the EAGLE simulations} 

\author[A. D. Hill et al.]{
Alexander D. Hill$^{1}$,\thanks{E-mail: a.d.hill@2017.ljmu.ac.uk (ADH)} Robert A. Crain,$^{1}$ Ian G. McCarthy$^{1}$ and Shaun T. Brown.$^{1}$
\\
$^{1}$Astrophysics Research Institute, Liverpool John Moores University, 146 Brownlow Hill, Liverpool L3 5RF, United Kingdom
}

\date{Accepted XXX. Received YYY; in original form ZZZ}

\pubyear{2021}

\begin{document}
\label{firstpage}
\pagerange{\pageref{firstpage}--\pageref{lastpage}}

\maketitle
\begin{abstract}
We present measurements of the intrinsic alignments (IAs) of the star-forming gas of galaxies in the EAGLE simulations. Radio continuum imaging of this gas enables cosmic shear measurements complementary to optical surveys. We measure the orientation of star-forming gas with respect to the direction to, and orientation of, neighbouring galaxies. Star-forming gas exhibits a preferentially radial orientation-direction alignment that is a decreasing function of galaxy pair separation, but remains significant to $\gtrsim 1\Mpc$ at $z=0$. The alignment is qualitatively similar to that exhibited by the stars, but is weaker at fixed separation. Pairs of galaxies hosted by more massive subhaloes exhibit stronger alignment at fixed separation, but the strong alignment of close pairs is dominated by ${\sim}L^\star$ galaxies and their satellites. At fixed comoving separation, the radial alignment is stronger at higher redshift. The orientation-orientation alignment is consistent with random at all separations, despite subhaloes exhibiting preferential parallel minor axis alignment. The weaker IA of star-forming gas than for stars stems from the former's tendency to be less well aligned with the dark matter structure of galaxies than the latter, and implies that the systematic uncertainty due to IA may be less severe in radio continuum weak lensing surveys than in optical counterparts. Alignment models equating the orientation of star-forming gas discs to that of stellar discs or the DM structure of host subhaloes will therefore overestimate the impact of IAs on radio continuum cosmic shear measurements. 
\end{abstract}


\section{Introduction}

The morphology, spin and orientation of galaxies are influenced by the tidal field of the cosmic large scale structure \citep[e.g.][]{Heavens88,Bond96,Wang18}. The coherence of the tidal field over large cosmic distances induces correlated orientations, a phenomenon often referred to as `intrinsic alignment' \citep[e.g.][]{heavens00, Croft00, Lee_Pen01, brown02, Jing, Mackey02, Aubert04}. This alignment represents a significant source of systematic uncertainty in cosmic shear measurements from weak lensing experiments, which aim to measure the distortion of the images of distant galaxies due to the lensing effect induced by intervening matter distribution along the line of sight. 

The observed correlation of the shapes of galaxies results from the \textit{apparent} alignment of their lensed images, and the \textit{intrinsic} alignment of their true orientations \citep[see][for a review]{troxel}. Much effort has been made to develop means of modelling intrinsic alignments in order to mitigate their impact on weak lensing surveys \citep[for reviews see][]{joachimi_review, Kiessling15, Kirk15}. Further motivation for modelling intrinsic alignment arises from its putative sensitivity to a diverse range of physical influences, such as the growth of angular momentum during galaxy formation \citep{Lee_Pen}, primordial gravitational waves \citep{chisari14}, modified gravity \citep{LHuillier} and self-interacting dark matter \citep{Harvey}. 

As the depth and fidelity of observations improves, commensurate improvements in the ability of weak lensing surveys to constrain cosmological parameters are increasingly limited by an incomplete understanding of the effect of baryons on the matter power spectrum and the intrinsic alignment of galaxies. \citet{amon21} argue that such uncertainties cost the Dark Energy Survey Year 3 \citep[DES Y3;][]{secco} cosmic shear measurements approximately two thirds of their constraining power. Intrinsic alignments have been estimated primarily using the analytic linear alignment model \citep{Catelan01, hirataA}, with the ansatz that the projected shapes of galaxies are linearly correlated with the projected tidal field. The linear alignment model accurately reproduces the inferred alignments of distantly-separated early-type galaxies ($\gtrsim 10h^{-1}~\mathrm{Mpc}$), however recent observations have shown it to underestimate the alignments of closer pairs \citep{singh, johnston19}. The non-linear alignment model \citep{bridle_king} makes use of the non-linear matter power spectrum while still assuming a linear response between galaxy shapes and the tidal field, and fares better at reproducing the observed alignments of galaxies at intermediate separations \citep{Hirata10}. It has thus enjoyed widespread adoption in the mitigation of intrinsic alignment uncertainty \citep[e.g.][]{joachimi2011, Heymans13, des_16}. Mitigating the uncertainty on shorter scales has motivated the development of more complex approaches, such as the quadratic alignment model \citep{Crittenden01}, perturbative expansions \citep{blazek, blazek15, Blazek19}, effective field theory \citep{vlah}, and applications of the halo model \citep{Schneider10, fortuna21}. 

In recent years, cosmological hydrodynamical simulations of the galaxy population, which simultaneously evolve dark matter (DM) and baryons, have achieved far better correspondence with the observed properties of the galaxy population than prior generations \citep[see e.g.][]{Somerville2015,Naab2017}. These simulations include treatments of the complex baryonic physics governing the formation and evolution of galaxies, which have been shown to impact the internal structure and the spatial distribution of haloes \citep[e.g.][]{Duffy10,schaller15a,Springel18}. Hydrodynamical simulations have been used to study the intrinsic alignment of galaxies even well within the non-linear, one-halo regime \citep{chisari15, Chisari16, codis15, tenneti15, vel15b, hilbert, Harvey, shi21}. They offer a means to obtain physical insights into the origins of galaxy shape correlations, and to assess the accuracy of analytic alignment models \citep{Samuroff}. 

Contemporary weak lensing experiments are dominated by optical/near-IR surveys, since to date only these have delivered imaging with the necessary source density required to extract a robust shear measurement. Successive data releases from the Kilo-Degree Survey (KiDS) have provided galaxy counts of ${\sim}10$ arcmin$^{-2}$ over 450 deg$^2$ and 1350 deg$^2$, respectively \citep{hildebrandt, heymans21}, while the DES Y3 dataset contains galaxy sources at 5.59 arcmin$^{-2}$ over 4143 deg$^2$ \citep{secco}. In principle, however, shear measurements can also be made using the extended radio continuum emission of the interstellar medium. To date, deep radio surveys such as the VLA-COSMOS \citep{Smolcic} and the SuperCLuster Assisted Shear Survey \citep[SuperCLASS;][]{battye, Harrison_2020, Manning} have yielded insufficient source counts of galaxies (${\lesssim}1$ arcmin$^{-2}$ over a few square degrees) to enable meaningful shear measurements, but surveys conducted with the forthcoming Square Kilometer Array (SKA) have the potential to yield competitive measurements. The first phase (SKA1) is forecast to achieve galaxy source counts of 2.27 arcmin$^{-2}$ over 5000 deg$^2$ \citep{ska_red}, while \citet{brown15} suggest that the most optimistic second phase (SKA2) implementation would deliver 30 galaxies arcmin$^{-2}$ over 3$\pi$ steradians. The two phases are forecast to provide cosmological constraining power on a par with Stage III and Stage IV optical surveys, respectively \citep{harrison}. 

Radio weak lensing surveys present numerous advantages: the characteristic redshift of sources will in general be higher than is the case for optical surveys, which due to increased mass along the line of sight will yield a stronger lensing signal, as well as enabling the analysis of the growth of structure at an earlier cosmic epoch \citep{brown15, camera, ska_red};  polarisation and/or kinematic information, available at no or little extra cost to the continuum observations, affords a means of mitigating against intrinsic alignment uncertainty by indicating the unlensed orientation \citep{blain,morales,deburghday,whittaker15}; the point spread function (PSF) of radio measurements is deterministic, enabling its precise removal, which is not the case for the PSF of optical observations \citep[e.g.][]{heymans12}; and there is the potential to measure the redshift distribution of sources directly from the radio observations via statistical detection of the (low signal-to-noise) 21-cm emission line \citep{Harrison_2017}. Arguably the chief benefit in conducting radio weak lensing surveys is the potential for cross-correlation with optical measurements, providing a means of mitigating the systematic measurement uncertainties afflicting each wavelength. The extended radio continuum emission is largely associated with star-forming gas, whose morphology and orientation need not be similar that of stellar component seen in the optical \citep[see e.g.][]{tunbridge}. Realisation of the potential of radio weak lensing surveys therefore requires accurate assessments of the intrinsic alignments of the star-forming gas component of galaxies.

In this study, we use the cosmological, hydrodynamical simulations of the EAGLE project \citep{schaye15, crain15} to measure the intrinsic alignments of the star-forming gas component of galaxies. These simulations self-consistently account for the back-reaction of baryons on the DM, and by modelling galaxies numerically need not appeal to geometric approximations for their size, morphology or orientation. EAGLE represents an ideal model on which to base this study, as the properties of the interstellar gas associated with its present-day galaxy population have been shown to correspond closely with observations \citep[see e.g.][]{lagos15, bahe16, crain17, dave20}, and it reproduces the `fundamental plane' of star formation \citep{lagos16}. This work builds on a prior study \citep[][hereafter \citetalias{hill21}]{hill21} in which we examined the morphology of star-forming gas distributions associated with EAGLE galaxies, and their internal alignment with their corresponding stellar and DM components. It also complements studies with EAGLE focussing on the alignments of the stellar component of galaxies \citep{vel15a,vel15b}. As per \citet{vel15b}, we focus on the orientation-direction and orientation-orientation intrinsic alignments in 3-dimensions and in projection. 

The paper is structured as follows. In Section~\ref{sec:methods} we briefly discuss details of the EAGLE simulation and outline our numerical methodology and sample selection criteria. In Section~\ref{sec:3d_results} we examine the intrinsic alignment of star-forming gas in 3-dimensions and assess its dependence on subhalo mass, redshift and its internal alignment with the DM of its host subhalo. In Section~\ref{sec:2d_results} we examine the intrinsic alignments in projection. In Section~\ref{sec:discussion} we discuss and summarise our findings. In the appendices we carefully assess the sensitivity of our results to the numerical resolution of the simulations, the details of the subgrid physics treatments directly governing the properties of star-forming gas, and the implementation of our shape and orientation characterisation method.

\section{Methods}
\label{sec:methods}

This section provides a brief overview of the EAGLE simulations (Section~\ref{sec:sims}) and introduces the methods used to identify haloes and galaxies (Section~\ref{sec:gals_haloes}), and to characterise their morphology and orientation (Section~\ref{sec:methods_shapefit}). Sample selection is discussed in Section~\ref{sec:sampling}, and the numerical characterisation of intrinsic alignments is discussed in Section~\ref{sec:IA}. 

\subsection{Simulations}
\label{sec:sims}

EAGLE \citep[Evolution and Assembly of GaLaxies and their Environments;][]{schaye15,crain15} is a suite of hydrodynamical simulations of the formation, assembly and evolution of galaxies within a $\Lambda$CDM cosmogony. The project's data has been publicly released, including  both raw snapshot data and processed data products such as galaxy and halo catalogues \citep{mcalpine16}. The simulations were evolved using a modified version of the $N$-body smoothed particle hydrodynamics (SPH) code \textsc{Gadget-3} \citep[last described by ][]{springel05_gad}, with the key modifications being a pressure-entropy implementation of SPH \citep{hopkins}, the use of the $C^2$ smoothing kernel \citep{wendland95}, a time-step limiter \citep{durier12}, and switches for artificial conduction \citep{price08} and viscosity \citep{cullen10}. The impact of these modifications on the simulated galaxy population is discussed by \citet{schaller15b}.

EAGLE implements sub-resolution (or `subgrid') routines to model physical processes that are not resolved numerically. The radiative cooling, heating and photoionisation of gas is treated element-by-element using the scheme of \citet{WSS_A}, assuming the presence of a spatially uniform, temporally evolving radiation field comprising the cosmic microwave background and the metagalactic ultra-violet background \citep[modelled by][]{HM01}. Per \citet{schaye_DV}, the interstellar medium (ISM) is treated as a single-phase fluid subject to a polytropic pressure floor, wherein gas particles denser than the metallicity-dependent threshold advocated by \citet{schaye04} are eligible for conversion into a stellar particle, with a probability proportional to the particle's star formation rate (SFR, itself a function of pressure), such that galaxies reproduce the observed Kennicutt-Schmidt relation \citep{kennicutt}.

Star particles are assumed to represent simple stellar populations with the \citet{chabrier03} initial mass function (IMF), which evolve and return mass to the ISM according to the model of \citet{wiersma09}, and inject feedback energy into their surroundings via stochastic isotropic thermal heating \citep{dallavecchia}. Black holes (BHs) of initial mass $10^5\Msun/h$ are seeded on-the-fly at the centres of haloes with mass greater than $10^{10}\Msun/h$, and grow via BH-BH mergers and Eddington-limited gas accretion at the Bondi-Hoyle rate, modulated by the circulation speed of gas local to the BH \citep{springel05_agn, rosas_guevara, schaye15}. Feedback energy released by this accretion, active galactic nuclei (AGN) feedback, is injected via stochastic isotropic thermal heating \citep{booth_schaye,schaye15}. The efficiency of stellar feedback and the BH accretion disc viscosity (which governs the modulation of the Bondi-Hoyle rate) were calibrated to reproduce the observed present-day galaxy stellar mass function (GSMF) and the sizes of disc galaxies, whilst the efficiency of AGN feedback was calibrated to reproduce the present-day relationship between galaxy stellar mass and BH mass. 

EAGLE adopts the \citet{planck} cosmological parameters,  $\Omega_{\mathrm{m}} = 0.307$,  $\Omega_{\mathrm{b}} = 0.04825$,  $\Omega_{\Lambda} = 0.693$, $\sigma_{\mathrm{8}} = 0.8288$, $n_{\mathrm{s}} = 0.9611$, $h = 0.6777$, $Y = 0.248$. The standard resolution simulations have particle masses corresponding to that of the flagship EAGLE simulation, Ref-L100N1504, from which our results are drawn. This is a periodic volume of side $L=100~\cMpc$\footnote{Throughout this paper we use the notation `c' and `p' to refer to comoving and proper units, respectively. At $z=0$ the distinction between proper and comoving units vanishes, so here the notation is dropped.} realised with $1504^3$ DM particles and an initially equal number of SPH particles, such that the initial gas and DM particle masses are $m_{\mathrm{g}}=1.81\times10^6~\mathrm{M_{\odot}}$ and $m_{\mathrm{DM}}=9.7\times10^6~\mathrm{M_{\odot}}$, respectively. The Plummer-equivalent gravitational softening length is fixed in comoving units to be 1/25 of the mean inter-particle separation, $\epsilon_{\mathrm{com}}=2.66~\ckpc$, limited to a maximum proper length of $\epsilon_{\mathrm{prop}}=0.7~\pkpc$. In Appendix~\ref{sec:app_convergence}, we test the numerical convergence behaviour of our results using a pair of high-resolution $L=25\cMpc$ simulations, with particle masses and softening scales smaller than those of Ref-L100N1504 by factors of 8 and 2, respectively.

The simulations thus marginally resolve the Jeans scales at the threshold density for star formation in the warm, diffuse phase of the ISM, but do not resolve the cold, molecular phase. The use of the aforementioned polytropic pressure law is needed to suppress the artificial fragmentation of star-forming gas, but a drawback of its use is the suppression of the formation of gas discs with a scale height less than the corresponding Jeans length \citep[see e.g.][]{trayford17}. \citetalias{hill21} examined the dependence of star-forming gas morphology on the normalisation of the pressure floor and found that reasonable variations induced systematic morphological changes that were small compared to the system-to-system scatter. We further examine the influence of the pressure floor, and that of the normalisation of the star-formation law, on the internal alignment of the various matter components of galaxies in Appendix~\ref{sec:app_subgrid}.

Another numerical limitation that can influence galaxy morphology is two-body scattering between stellar and DM particles of unequal mass, which can also lead to artificial heating of the stellar component  \citep{ludlow}. We therefore caution that discs of gas and stars are both generally thicker in EAGLE than in real galaxies. Whilst unlikely to impact galaxy orientations, these limitations may affect measures dependent upon galaxy morphology, such $\epsilon_{\mathrm{g+}}$ and $\epsilon_{++}$ (Section~\ref{sec:ia_measure_2D}).

\subsection{Identification of galaxies and haloes}
\label{sec:gals_haloes}

The friends-of-friends (FoF) algorithm is used to identify haloes within the DM distribution, using a linking length one-fifth of the mean interparticle separation. Particles of other types are assigned to the group, if any, of their closest DM particle neighbour. The \textsc{subfind} algorithm is used to find overdensities within the FoF haloes, identifying peaks separated by saddle points in the density distribution \citep{springel01, dolag09}. These overdensities are termed `subhaloes', further labelled as centrals if they contain the particle with the lowest gravitational potential energy, and satellites otherwise. Galaxies are defined as the baryonic component of subhaloes. The position of a galaxy (and its subhalo) is defined by that of its particle with the lowest gravitational potential energy. Subhalo properties are computed by aggregating the relevant properties of their constituent particles.

\subsection{Characterisation of the morphology and orientation of galaxy components}
\label{sec:methods_shapefit}

The shapes and orientations of galaxies and their subhaloes are quantitatively characterised by fitting a 3-dimensional ellipsoid to the relevant particle distribution. This ellipsoid is characterised by major, intermediate and minor axis lengths ($a, b, c$) and vectors ($\vec{e_1}, \vec{e_2}, \vec{e_3}$). The characteristics of the ellipsoid are computed via the mass distribution tensor of the relevant particles:
\begin{equation} 
M_{ij}= \frac{\sum_{p} w_p r_{p,i} r_{p,j}}{\sum_{p} w_p},
\label{eq:inertiatensor}
\end{equation}
where the sum is over all particles, $p$. Here $r_{p,i}$ is the $i$-th element of a particle $p$'s coordinate vector relative to the galaxy centre, and $w_p$ is a weighting factor. As the mass distribution tensor and the inertia tensor of an object share common eigenvalues and eigenvectors, it is common to use the two terms interchangeably. In this work we will refer to the inertia tensor.

The choice of the inertia tensor is not unique \citep[see e.g.][]{bett12}. The simplest form weights particles by their mass \citep[$w_p = m_p$, e.g.][]{davisB, cole} and we adopt this approach when considering the stellar and DM components of haloes. To mimic the structure of radio continuum-luminous regions, whose luminosity broadly correlates linearly with the local SFR \citep[see e.g.][]{condon92, schober}, we consider gas particles but weight them by their SFR rather than their mass ($w_p = \dot{m}_{p,\star}$). The SFR of gas particles is precisely zero unless the particle is both denser than the metallicity-dependent star formation threshold, and has a temperature within $0.5\,{\rm dex}$ of the polytropic pressure floor. We do not consider radio continuum emission due to AGN, since this is not extended.

As per \citetalias{hill21}, we use an iterative form of the reduced inertia tensor \citep[see also][]{dubinski_carlberg, bett12, schneider, thob19}. The reduced form of the tensor suppresses the influence of structures in the subhalo outskirts by down-weighting the contribution of particles at large (ellipsoidal) radii (i.e. $w_p = \dot{m}_{p,\star}/\Tilde{r}^{2}_{p}$, where $\Tilde{r}_{p}$ is the ellipsoidal radius of the particle), whilst the iterative scheme enables the best-fit ellipsoid to adapt to particle distributions that are markedly different in morphology to the initial particle selection. Since the simplest choice for the latter is a sphere (or, in 2-dimensions, a circle), the iterative tensor is advantageous when characterising flattened systems such as galaxy discs. The best fit ellipsoid is therefore first computed within a spherical aperture of radius $r_{\mathrm{ap}} = 30~\mathrm{pkpc}$, where this value is chosen for consistency with that commonly used when computing galaxy properties by aggregating particle properties \citep[see e.g. Section 5.1.1 of][]{schaye15}. In 2-dimensions, a circular aperture of $r_{\mathrm{ap}}=\mathrm{max}(30~\mathrm{pkpc},2r_{1/2,\mathrm{SF\mbox{-}Gas}})$ is used, where $2r_{1/2,\mathrm{SF\mbox{-}Gas}}$ is the half-mass radius of the star-forming gas. It is then re-computed iteratively, using the particles enclosed by the best-fit ellipsoid of the previous iteration. Complete details of the algorithm are given in Section~2.3 of \citetalias{hill21}. In Appendix~\ref{sec:app_tensor}, we assess the sensitivity of intrinsic alignments to the chosen form of the inertia tensor, and show that it has a milder influence on the intrinsic alignments inferred for the star-forming gas than is the case for the stellar and DM components of subhaloes. 

\subsection{Sample selection}
\label{sec:sampling}

Unless otherwise stated, we adopt the same sampling criteria used by \citetalias{hill21}. We require at least 100 particles each of star-forming gas, stars and DM to be present within the final converged ellipsoid. This threshold is motivated by numerical tests (see Appendix~A of \citetalias{hill21}), which indicate that a minimum of 100 particles is needed to recover the sphericity of particle distributions with a measurement error of less than 10 percent. We further require that the star-forming gas distribution is reasonably axisymmetric, since we fit it with an axisymmetric shape. We use an adapted form of the axisymmetry measure introduced by \citet{trayford19}, $A_{\mathrm{3D}}$, whereby we bin the mass of star-forming gas particles in pixels of solid angle about the galaxy centre using \textsc{Healpix} \citep{gorski05}, and compute the fractional difference in the mass of opposing pixels. For inclusion in our fiducial sample, galaxies must have $A_{\mathrm{3D}} \leq 0.6$. 

At $z=0$, both the particle sampling and axisymmetry criteria are satisfied by 6764 galaxies. The particle sampling criteria in particular introduce a strong selection bias, especially at low subhalo mass since they correspond to a minimum stellar mass of ${\sim}10^8\Msun$ and a minimum SFR of $\simeq 6\times 10^{-2}~\mathrm{M_{\odot}yr^{-1}}$. Our sample includes approximately (0.1, 10, 80) percent of all subhaloes of total mass $\log_{10} (M_{\mathrm{sub}}/\Msun) \sim (10,11,12)$, and (16, 65, 60) percent of all subhaloes of stellar mass $\log_{10} (M_{\mathrm{\star}}/\Msun) \sim (9,10,11)$. 

\subsection{Intrinsic alignments}
\label{sec:IA}

Cosmic shear, the correlation in the shapes of distant galaxies whose images have been distorted by the lensing effect of the large scale structure of the Universe, is detectable only in the correlation of the shapes of many background galaxies. In the limit of weak gravitational lensing, the observed ellipticity ($e^{\mathrm{obs}}$) of a galaxy may be expressed as the sum of its intrinsic shape ($e^{\mathrm{int}}$) and the shear distortion due to lensing ($\gamma$)
\begin{equation}
    e^{\mathrm{obs}} = e^{\mathrm{int}}+\gamma.
\end{equation}
In the absence of intrinsic alignment, $\langle e^{\mathrm{int}}\rangle=0$. Therefore for a sufficiently large sample of galaxies in a given patch of sky, any non-zero measurement of $e^{\mathrm{obs}}$ may be interpreted as a measurement of the shear due to the influence of the integrated mass density along the line of sight.

In practice non-random galaxy alignments are a significant source of systematic bias. The projected two-point correlation function between the shapes of galaxies is defined as
\begin{equation}
\begin{aligned}
    \langle e^{\mathrm{obs}} e^{\mathrm{obs}}\rangle &=  
    \langle\gamma\gamma\rangle +
    \langle\gamma e^{\mathrm{int}}\rangle +
    \langle e^{\mathrm{int}}\gamma\rangle +
    \langle e^{\mathrm{int}} e^{\mathrm{int}}\rangle.\label{eq:II_GG}
\end{aligned}
\end{equation}
The right hand side of this equation may also be expressed as GG + GI + IG + II. GG is the shear-shear auto-correlation term, and it encapsulates the correlation caused by the mutual lensing of galaxy images by some common intervening matter distribution. II is the intrinsic-intrinsic auto-correlation, caused by a close pair of galaxies being mutually aligned due to their independent alignment with some common large-scale structure. The shear-intrinsic cross-correlation term GI is caused by cases where the observed shapes of two galaxies ($g_i$, $g_j$) that reside at different redshifts ($z_i < z_j$) are correlated due to a massive object at $\simeq z_i$ acting as both a lens for $g_j$ and a source of intrinsic alignment for $g_i$. The mechanism causing the IG term is the similar to GI, except here the massive object resides at $\simeq z_j$. In practice $\mathrm{IG} = 0$, as a background object cannot lens a foreground galaxy.

The observed intrinsic alignment of galaxies in projection is caused primarily by their true 3-dimensional orientation and morphology. In this paper we explore both the 2- and 3-dimensional intrinsic alignments of galaxies in order to investigate both their expected impact on radio cosmic shear measurements, and to determine their physical cause. We largely refer to `orientation-orientation' and `orientation-direction' alignments, where the former concerns the orientations of a pair of galaxies, and the latter compares the orientation of one galaxy with the direction vector connecting it with a neighbour. Orientation-orientation alignment is straightforwardly the II term. Orientation-direction alignment concerns the preference for a galaxy to be orientated with respect to the location of another galaxy, and hence by extension the ambient large-scale structure. Orientation-direction alignment is therefore related to the GI term. \citet{joachimi2011} provides a derivation of the GI power spectrum from the ellipticity correlation function. In what follows, we use the term `intrinsic alignments' to refer to both the II and GI terms.

\subsubsection{Measuring intrinsic alignments in 3-dimensions}
\label{sec:ia_measure_3D}

\begin{figure}
\centering
\hspace{-0.2cm}
     \includegraphics[width = 0.46\textwidth]{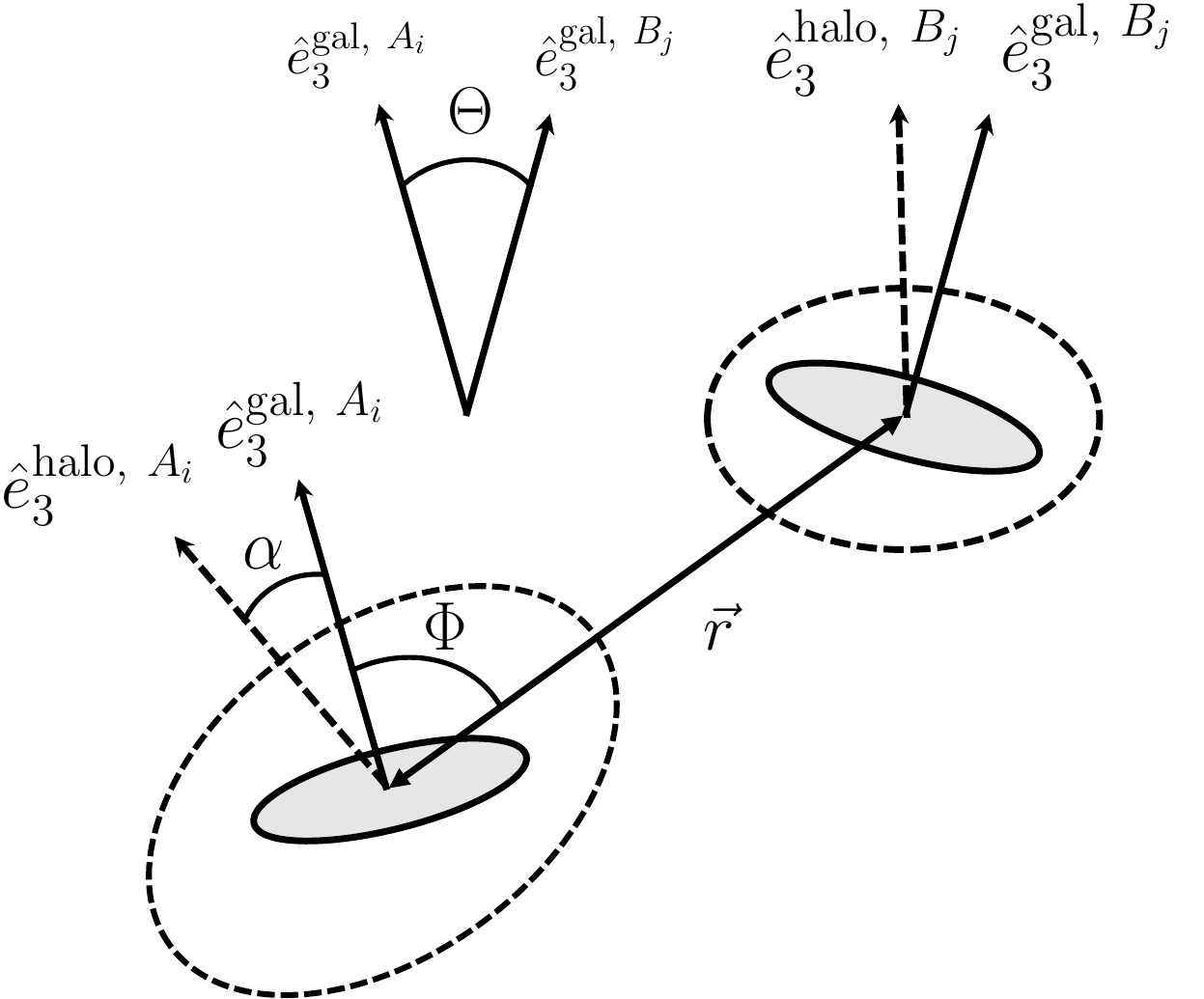}
\caption{Schematic representation of the 3-dimensional orientation-direction and orientation-orientation intrinsic alignments. The centres of subhaloes $A_i$ and $B_j$ are separated by distance $|\vec{r}|$. The orientation of the galaxy (grey shaded ellipsoid) is misaligned with respect to the orientation of its dark matter subhalo (dashed ellipsoid) by the angle $\alpha = \cos^{-1}(|\hat{e}_{3}^{\mathrm{gal}, A_i}\cdot \hat{e}_{3}^{\mathrm{halo}, A_i}|)$. The orientation-direction alignment of galaxies is defined as $\Phi = \cos^{-1}(|\hat{e}^{\mathrm{gal}, A_i}_{3} \cdot \hat{r}|)$, while the orientation-orientation alignment is $\Theta = \cos^{-1}(|\hat{e}_{3}^{\mathrm{gal}, A_i} \cdot \hat{e}_{3}^{\mathrm{gal}, B_j}|)$.}
\label{fig:diagram}
\end{figure}

To measure intrinsic alignments we require the position and orientation of a pair of galaxies, necessitating two samples: $\mathcal{A} = \{A_{1}, A_{2}, ..., A_{n}\}$ and $\mathcal{B} = \{B_{1}, B_{2}, ..., B_{m}\}$. Fiducially, $\mathcal{A}$ and $\mathcal{B}$ are both the complete sample of 6764 galaxies that satisfy the criteria outlined in Section~\ref{sec:sampling}. To assess the sensitivity of alignments to various galaxy properties, we further sub-sample $\mathcal{A}$ and/or $\mathcal{B}$. For example, if we wish determine the orientation-direction alignment between galaxies of different subhalo masses, we sub-sample $\mathcal{A}$ and $\mathcal{B}$ accordingly and indicate this in figures with the notation $(\mathcal{A})[M^A_{\mathrm{low}},M^A_{\mathrm{high}}]$ and $(\mathcal{B})[M^B_{\mathrm{low}},M^B_{\mathrm{high}}]$. Sub-sampling by other properties to assess different dependencies is similarly indicated.

A graphical depiction of the 3-dimensional intrinsic and internal alignments explored in this paper is shown in Fig.~\ref{fig:diagram}. The 3-dimensional alignments are defined as the cosine of an angle of interest for a galaxy pair separated by some vector $\vec{r}$, $\cos(\chi)$. To assess the influence of galaxy separation, we compute the mean of $\cos(\chi)$ in bins of galaxy pair separation. A pair is comprised of one galaxy from $\mathcal{A}$ and one from $\mathcal{B}$, such that the number of galaxy pairs $N_{\mathrm{p}}=n\times m$. In the case of the orientation-direction alignment of a galaxy pair ($A_{i}$, $B_{j}$) with positions ($\vec{x}$, $\vec{x} + \vec{r}$), respectively, we measure the angle between $A_{i}$'s morphological minor axis, $\vec{e_3}$, and the direction vector connecting the positions of the pair, $\vec{r}$, such that 
\begin{equation}
    \cos(\Phi(r))= (|\hat{e}_{3}^{A_i} \cdot \hat{r}|),\label{eq:or_dir}
\end{equation} 
where carets denote unit vectors. Note that taking the absolute value of the vector dot product bounds $\Phi$ between 0 and $\pi/2$, and hence $\cos(\Phi)$ between 0 and 1. The expectation value of $\cos(\Phi)$ for a random distribution of vectors in 3-dimensions is 0.5, $\cos(\Phi)=1$ indicates perfect alignment between the two vectors, while $\cos(\Phi)=0$ indicates perfect anti-alignment. Since we measure $\Phi$ with respect to the morphological minor axis, radial alignment (the preference for the disc plane to be aligned with the direction to a neighbour) is signified by $\cos(\Phi)<0.5$.

In the case of the orientation-orientation alignment, we compare the orientations of both $A_i$ and $B_j$ as
\begin{equation}
    \cos(\Theta(r))= (|\hat{e}_{3}^{A_i} \cdot \hat{e}_{3}^{B_j}|).\label{eq:or_or}
\end{equation}
The expectation value of $\cos(\Theta)$ for a random distribution of 3-vectors is again 0.5, with $\cos(\Theta)=1$ indicating that the minor axes of two galaxies are exactly parallel, and $\cos(\Theta)=0$ that they are exactly perpendicular.

\begin{figure}
\centering
\hspace{-0.2cm}
     \includegraphics[width = 0.48\textwidth]{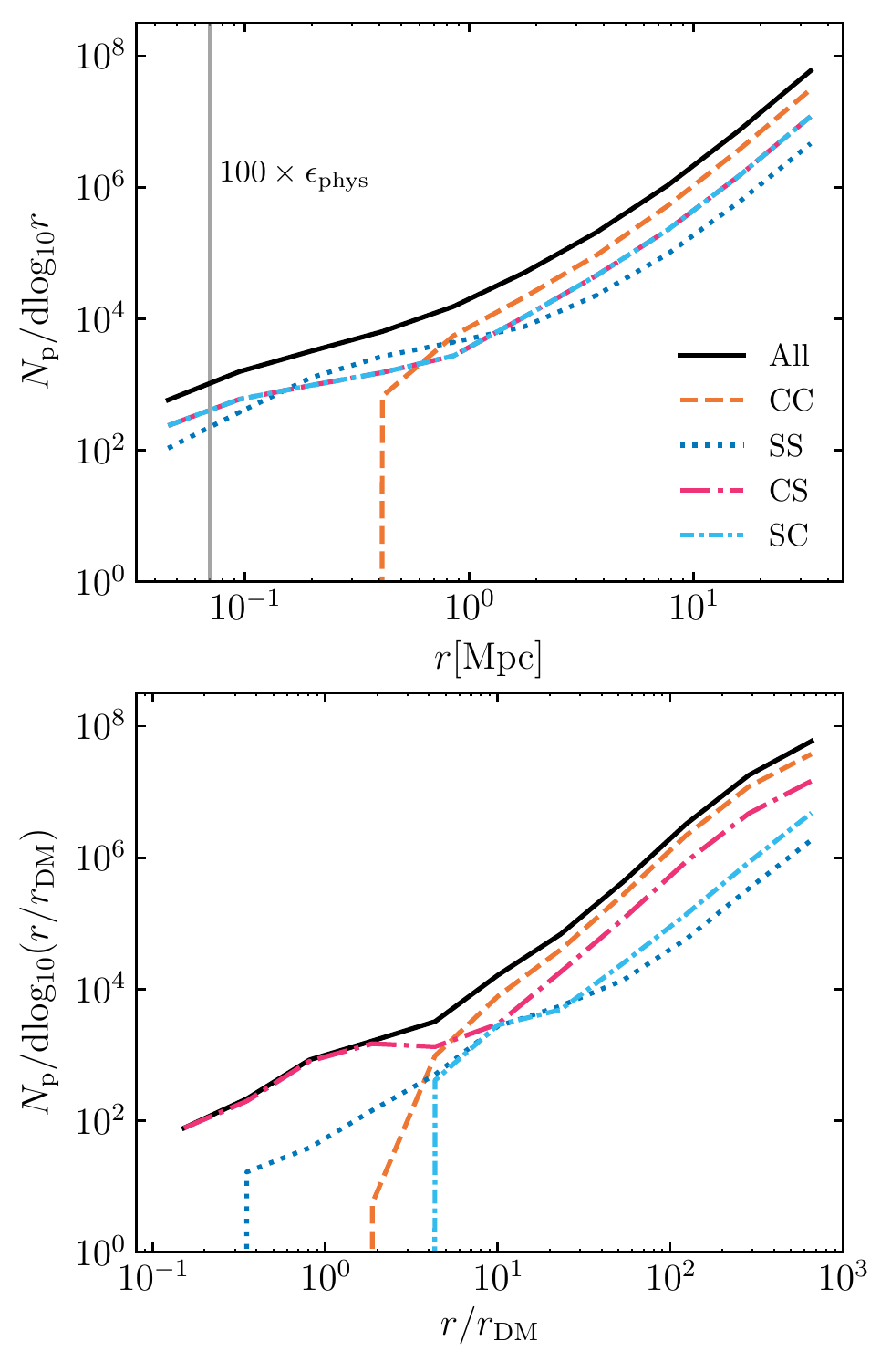}
\caption{Galaxy pair counts, $N_{\rm p}$, as a function of separation for our fiducial sample. Counts are shown for the entire sample (black curve) and separately for the contributions of various pairings of central (C) and satellite (S) galaxies (see legend). Pair counts are shown as a function of absolute separation in the upper panel, and as a function of separation normalised by the dark matter half-mass radius of the primary galaxy's subhalo in the lower panel. The vertical grey line in the top panel is drawn at $100\epsilon_{\mathrm{phys}}$, where $\epsilon_{\mathrm{phys}} = 0.70\pkpc$ is the maximum proper softening length of the Ref-L100N1504 simulation.}
\label{fig:censat_pairs}
\end{figure}

We examine alignments as a function of both the absolute 3-dimensional distance between galaxies, and the distance normalised by the half-mass radius of the DM distribution ($r/r_{\mathrm{DM}}$) of the primary galaxy of each pair. We only consider separations less than half of the simulation boxsize. 

\begin{figure*}
\centering
\hspace{-0.2cm}
     \includegraphics[width = 0.95\textwidth]{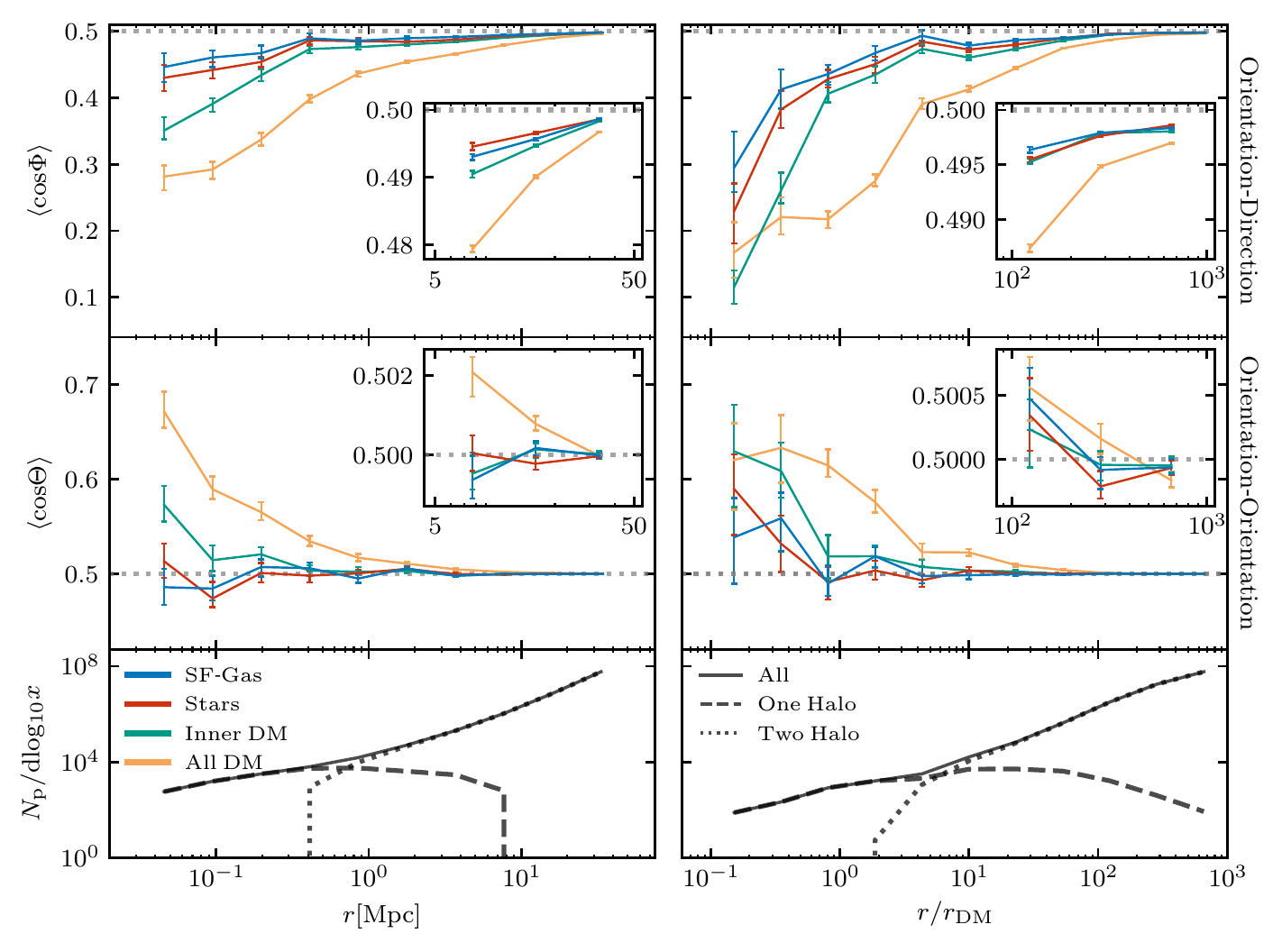}
\caption{The present-day orientation-direction (top row) and orientation-orientation (middle row) intrinsic alignments as a function of galaxy pair separation for the star-forming gas (blue curves), stars (red) and DM (inner subhalo: green, entire subhalo: yellow) of our fiducial sample. Dotted horizontal lines correspond to the expectation value for randomly-orientated 3-vectors (i.e. no intrinsic alignment). Inset panels zoom in to highlight the small but statistically significant intrinsic alignments at large-separations. The bottom row shows the corresponding pair counts (solid curve), and the contributions of subhaloes sharing the same FoF halo (one-halo term: dashed) and those in different haloes (two-halo; dotted). The left and right columns correspond, respectively, to the separation in absolute terms and that normalised by the DM half-mass radius of the primary galaxy's subhalo ($r_{\mathrm{DM}}$). Error bars denote the bootstrap-estimated uncertainty on the measurements. Curves are drawn only for bins sampled by at least 10 galaxies. The orientation-direction alignment increases at decreased separation, and is weaker for the star-forming gas than the other matter components. No significant orientation-orientation alignment is seen for the star-forming gas.}
\label{fig:matter_dep}
\end{figure*}

Fig.~\ref{fig:censat_pairs} shows the number of galaxy pairs constructed from our fiducial sample as a function of their separation, both in terms of absolute distance (top panel) and that normalised by the half-mass radius of the primary subhalo's dark matter distribution, $r_{\mathrm{DM}}$ (bottom panel). For context, a grey vertical line is drawn at 100 times the maximum proper softening length, $\epsilon_{\mathrm{phys}}=0.7\pkpc$. The plot also shows the relative contribution of various combinations of central (C) and satellite (S) pairings, for example CS denotes a pairing where $A_i$ is a central and $B_j$ is a satellite. CC pairings are the dominant contributor to the overall pair counts, and hence the intrinsic alignments, at distant separations in both absolute ($r>1~\mathrm{Mpc}$) and halo-normalised terms ($r/r_{\mathrm{DM}} >10$). At $r/r_{\mathrm{DM}} <1$, galaxy pairings are entirely contained within one halo, with CS comprising the majority of pairings and SS making only a small contribution. Note that the CS and SC counts are identical when binned by absolute separation, but not when binned by $r/r_{\mathrm{DM}}$ since the primary halo differs in each case. Moreover, CS, SC and SS pairings do not necessarily reside in the same FoF halo, hence the non-vanishing contribution at large separation. At short separations it is however generally the case that such pairings do share a FoF halo (see the bottom panels of Fig.~\ref{fig:matter_dep}).  

We estimate the uncertainty on the alignment measurements using bootstrap re-sampling \citep[e.g.][]{Barrow}. Within each radial bin containing $N_{\mathrm{p}}$ galaxy pairs, we randomly select with replacement $N_{\mathrm{p}}$ pairs and recompute $\langle\cos\chi\rangle$. This is repeated 100 times, and we show the 16$^{\mathrm{th}}$ and 84$^{\mathrm{th}}$ percentiles of this distribution of measurements on plots with error bars. As detailed in Appendix~\ref{sec:app_sampling}, we also estimate the measurement uncertainty stemming from the finite size of the simulation volume, which limits both the number of pairs we are able to sample at each separation, and the diversity of the environments from which they are drawn. We approximate the fractional uncertainty as a function of $N_{\mathrm{p}}$ using the power law function $f(N_{\mathrm{p}}) = AN_{\mathrm{p}}^{k}$, with $A = 65.7$ (-65.2) and $k = -0.524$ (-0.523) for upper and lower bounds, respectively.

In Section \ref{sec:internal_alignment} we assess the impact that the internal alignment between a galaxy's star-forming gas and the DM distribution of its host subhalo has on the intrinsic alignments of galaxies. This internal alignment is characterised by the `misalignment angle'
\begin{equation}
    \alpha = \cos^{-1}(|\hat{e}_{3}^{\mathrm{gal}}\cdot \hat{e}_{3}^{\mathrm{halo}}|),\label{eq:alpha}
\end{equation}
where $\hat{e}_{3}^{\mathrm{gal}}$ and $\hat{e}_{3}^{\mathrm{halo}}$ are the unit vectors parallel to the minor axis of the galaxy's star-forming gas and that of its DM, respectively. Since the internal alignments of components can exhibit a significant radial variation \citep[see e.g.][]{vel15b}, we compute the misalignment angle with respect to $\hat{e}_{3}^{\mathrm{halo}}$ in two ways, the first applying to the DM the same initial $30\pkpc$ aperture that is used for the star-forming gas when computing the inertia tensor, and the second considering all DM particles bound to the subhalo. We refer to these misalignment angles as $\alpha_{\mathrm{in}}$ and $\alpha_{\mathrm{all}}$, respectively.

\subsubsection{Measuring intrinsic alignments in 2-dimensions}
\label{sec:ia_measure_2D}

The projected morphology of a galaxy depends on both its intrinsic 3-dimensional morphology and its orientation with respect to the observer. Weak lensing studies typically approximate the morphology of galaxy `images' as a simple ellipse\footnote{The procedure of fitting this shape is however far from simple, see e.g. \citet{kaiser_squires} or \citet{zuntz}.}, characterised by the ratio of its axis lengths and its orientation. We therefore approximate the image morphology of simulated galaxies by fitting ellipses to their particle distributions following projection along one of the Cartesian axes of the simulation volume, using the 2-dimensional form of the reduced inertia tensor. In \citetalias[][]{hill21} we showed that, since the star-forming gas distribution of galaxies is typically more flattened than is the case for its stellar component, there is greater variance in the projected ellipticity (i.e. `shape noise') of the radio continuum image than the optical image. 

The projected galaxy morphology is commonly described by the complex ellipticity \citep[e.g.][]{Mandelbaum06}, with components given by 
\begin{equation}
    (e_{+}, e_{\times}) = \frac{b^2-a^2}{b^2+a^2}[\cos(2\phi), \mathrm{sin}(2\phi)],
\end{equation}
where $\phi$ is the orientation angle\footnote{Note that $\Phi$ and $\phi$ correspond to the orientation-direction alignment angle in 3- and 2-dimensions, respectively.}, and $a$ and $b$ are the minor and major axis lengths, respectively. In contrast to the 3-dimensional morphology, there is no reason to prefer the use of the minor axis to define the image orientation, so we follow convention and define $\phi$ as the angle subtended by the major axis of a galaxy $A_i$ and some tracer of the density distribution, in this case a galaxy from the $\mathcal{B}$ sample\footnote{In the literature it is common that a galaxy pair is described as belonging to a shape ($S$) and density sample ($D$), particularly when relating to the construction of Landy-Szalay estimator.}. $e_{+}$ is the radial component of the ellipticity, $e_{\times}$ is the 45$^\circ$-rotated component. The `total' (orientation-free) ellipticity is specified by $e = \sqrt{e_{+}^2 + e_{\times}^2}$.

We characterise the projected orientation-direction intrinsic alignment as a function of projected separation following
\begin{equation}
        \epsilon_{\mathrm{g+}}(r_{\mathrm{p}}) = \sum_{i\neq j | r_{\mathrm{p}}}^{N_\mathrm{p}}\frac{e_{+}(i|j)}{N_{\mathrm{p}}},
        \label{eq:eg_plus}
\end{equation}
which is also known as the average intrinsic shear of galaxies \citep[e.g.][]{singh}, and the projected orientation-orientation intrinsic alignment is computed as
\begin{equation}
    \epsilon_{++} = \sum_{i\neq j|r_{p}}^{N_\mathrm{p}} \frac{e_{+}(j|i) e_{+}(i|j)} {N_{\mathrm{p}}}.
    \label{eq:epp}
\end{equation}
We also present the average projected orientation-direction alignment angle, computed using the estimator
\begin{equation}
    \langle \phi \rangle = \sum_{i\neq j|r_{p}}^{N_\mathrm{p}}\frac{\phi}{N_{\mathrm{p}}},\label{eq:phi}
\end{equation}
This measure provides a more intuitive view of the projected alignments as a function of separation.

We consider only pairs separated along the projection axis by less than $4\pMpc$ in order to restrict our analyses to galaxies sharing similar large-scale structure, however we find that our results are relatively insensitive to plausible choices of this value. For the avoidance of confusion, we follow \citet{Mandelbaum06} and remark that positive values of $e_{+}$ indicate radial alignment, i.e. a tendency for the major axis of galaxies to point towards overdense regions of galaxies, which is the opposite of the often-applied convention in the weak lensing lensing literature that a positive shear signal corresponds to tangential alignment. 

\section{Intrinsic Alignments In 3-Dimensions}
\label{sec:3d_results}

In this section we examine the 3-dimensional intrinsic alignments of the star-forming gas of galaxies. In Section~\ref{sec:3D_matter_diff} we compare the present-day orientation-direction and orientation-orientation alignments, and compare with the analogous alignments exhibited by galaxies' stars and DM. In Section~\ref{sec:mass_dep} we assess the dependence of the alignments on subhalo mass, in Section~\ref{sec:redshift_dep} we explore their evolution with redshift, and in Section~\ref{sec:internal_alignment} we assess the sensitivity of the orientation-direction alignment to the internal alignment of star-forming gas with the DM distribution of its host subhalo.

\subsection{Intrinsic alignments of star-forming gas, stars and dark matter}
\label{sec:3D_matter_diff}

\begin{figure*}
\centering
\hspace{-0.2cm}
     \includegraphics[width = 1.\textwidth]{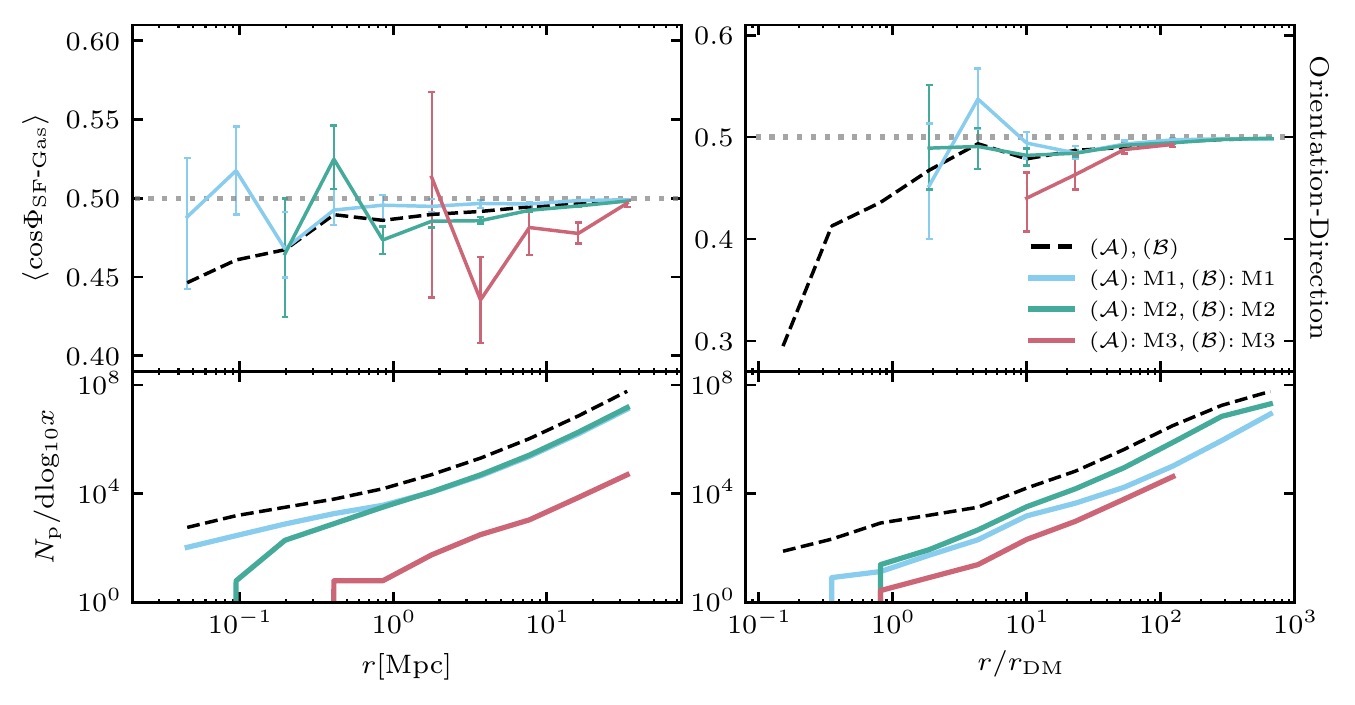}
\caption{The present-day orientation-direction alignment of the star-forming gas component of galaxy pairs of similar subhalo mass, as a function of pair separation. The M2 bin (green curves) includes galaxies with dynamical mass broadly similar to that of the Milky Way ($\mathrm{log_{10}M_{\mathrm{sub}}}/M_{\mathrm{\odot}} \in (11.47, 12.77)$), while the M1 (blue) and M3 (red) bins include subhaloes of mass below and above this range, respectively. Dashed black curves corresponds to the $\mathcal{A}$ and $\mathcal{B}$ samples without mass binning (i.e. the blue curves from Fig.~\ref{fig:matter_dep}). Dotted horizontal lines correspond to the expectation value for randomly-orientated 3-dimensional vectors (i.e. no intrinsic alignment). The bottom row shows the corresponding pair counts. The left and right columns correspond, respectively, to the separation in absolute terms and that normalised by the DM half-mass radius of the primary galaxy's subhalo ($r_{\mathrm{DM}}$). Error bars denote the bootstrap-estimated uncertainty on the measurements. Curves are drawn only for bins sampled by at least 10 galaxies. Orientation-direction alignment increases with subhalo mass for well-sampled separation bins. Normalising distances by $r_{\mathrm{DM}}$ reduces, but does not eliminates, the mass dependence.}
\label{fig:mass_dep_auto}
\end{figure*}

In Fig~\ref{fig:matter_dep}, we show the mean orientation-direction ($\langle \cos\Phi \rangle$, top row) and orientation-orientation ($\langle \cos\Theta \rangle$, middle row) alignments as a function of separation for the star-forming gas (blue curves), stars (red) and DM within the subhaloes of our sample. As noted in Section \ref{sec:ia_measure_3D} we consider the DM bound to the entire subhalo (yellow) and that within the inner regions (green). Dotted horizontal lines correspond to the expectation value for randomly-orientated 3-vectors (i.e. no intrinsic alignment). Inset panels zoom-in to highlight the small but significant intrinsic alignments at large-separations. The bottom row shows the total number of galaxy pairs (solid curve) as a function of separation, with the dashed and dotted curves denoting the contributions of galaxies sharing the same FoF halo (one-halo term) and those in different FoF haloes (two-halo term), respectively. At $z=0$ galaxy pairs in our sample with separations $\lesssim 0.8~\mathrm{Mpc}$ typically reside within the same FoF halo, whilst at $r > 1~\mathrm{Mpc}$ pairs typically belong to different FoF haloes. The tail of one-halo pairs towards large values of $r/r_{\mathrm{DM}}$ is due to pairs where the primary galaxy is a satellite.

The star-forming gas of galaxies exhibits a non-random orientation-direction alignment out to large separations (10s of $\mathrm{Mpc}$), with $\langle \cos\Phi \rangle$ decreasing farther below $0.5$ (the expectation value in the absence of intrinsic alignment) towards shorter separations. Therefore, as has been widely shown for the stellar component of simulated galaxies \citep[e.g.][]{chisari15, Chisari16, tenneti15, vel15b, Harvey}, the star-forming gas component exhibits a tendency to orient in a systematic fashion with respect to the ambient large-scale structure, with relatively close pairs being preferentially radially aligned. However, at all separations, the alignment is weaker than is the case for the stars, and increasingly so for the inner DM and entire DM distributions in turn. At $r = 10~\mathrm{Mpc}$, where the two-halo term is dominant, $\langle \cos\Phi \rangle = (0.495, 0.494, 0.492, 0.482)$\footnote{Values quoted are computed via a linear interpolation between the two closest known points.} for star-forming gas, stars, the inner DM halo and the entire DM halo, respectively. At $r = 1~\mathrm{Mpc}$, approximately the scale of the one- to two-halo transition, the corresponding values are $(0.487, 0.485, 0.477, 0.439)$, and at $r = 0.1~\mathrm{Mpc}$, a scale for which the one-halo term dominates, $\langle \cos\Phi \rangle = (0.461, 0.443, 0.393, 0.295)$. At all sampled separations (the upper end of which is limited by the simulation boxsize) and for all matter components, the deviation from random is significantly larger than the estimated uncertainty on the measurement, indicating that $\langle \cos\Phi \rangle$ is inconsistent with a random distribution of alignments.

As is clear from the right-hand column, the orientation-direction alignment is particularly strong within a few $r_{\mathrm{DM}}$. Pairs in this regime generally share the same FoF halo, which dominates the local environment. Nevertheless, significant intrinsic alignment of the star-forming gas persists to $r \sim 10^2 r_\mathrm{DM}$. Considering entire DM haloes, strong alignments persist beyond $r \sim 10^2 r_\mathrm{DM}$, where central-central pairings are the largest contributors to the pair counts.

At fixed separation all matter components exhibit an orientation-orientation alignment that is much weaker than the corresponding orientation-direction alignment. Binned by absolute separation, the star-forming gas components of neighbouring galaxies exhibit no significant non-random alignment, but binning by $r/r_{\mathrm{DM}}$ reveals a small but significant intrinsic alignment at short separations ($r \lesssim r_{\mathrm{DM}}$). The alignment here is primarily due to galaxies that share a common FoF halo, and the tendency for $\langle\cos\Theta\rangle > 0.5$ indicates a preference for their minor axes to be parallel. The stellar component exhibits similar behaviour, with orientations broadly consistent with a random distribution when binned by absolute separation, but a significant intrinsic alignment is apparent at $r \lesssim r_{\mathrm{DM}}$. We note that \citet{vel15b} examined the orientation-orientation alignment of the stellar component of ${\sim}L^\star$ galaxies in the Ref-L100N1504 simulation, and found similarly weak (or absent) alignment when using a similar aperture to that we use here (see their Fig. 4). As per the orientation-direction case, the subhalo DM component exhibits much stronger orientation-orientation alignment at fixed separation than the baryonic components, such that a significant parallel alignment ($\langle \cos\Theta\rangle > 0.5)$) persists to separations of ${\sim}10~\mathrm{Mpc}$, or $r \sim 10 r_{\mathrm{DM}}$, when one considers subhaloes in their entirety. In this case, at $r = (10, 1, 0.1)~\mathrm{Mpc}$ we find $\langle \cos\Theta \rangle = (0.502, 0.516, 0.588$), respectively. The alignment of the inner regions of subhaloes is weaker, but still much stronger than that of the baryonic components, for example at $r = 0.1~\mathrm{Mpc}$ we find $\langle \cos\Theta \rangle= 0.515$.

\subsection{Intrinsic alignment of star-forming gas as a function of mass}
\label{sec:mass_dep}

We next examine the influence of subhalo mass on the intrinsic alignments of star-forming gas distributions. Since the orientation-orientation alignment is effectively consistent with random except for close central-satellite pairs, we consider only the orientation-direction alignment.

We consider three subhalo mass bins, with the intermediate bin ($\mathrm{M2}$) corresponding approximately to the dynamical mass of the Milky Way, $\mathrm{log_{10}M_{\mathrm{sub}}}/M_{\mathrm{\odot}}\in (11.47, 12.77)$. The bins $\mathrm{M1}$ and $\mathrm{M3}$ contain all subhaloes from the fiducial sample with mass less than and greater than the range spanned by $\mathrm{M2}$, respectively. We employ this binning scheme to enable a more straightforward comparison with \citet{vel15b}, who used the M2 mass bin when considering intrinsic alignments of galaxies within the Ref-L100N1504 simulation. The completeness of these mass-selected sub-samples with respect to the entire subhalo population of this simulation is complicated by the selection criteria of the fiducial sample, particularly the requirements for 100 star-forming gas particles and axisymmetry, since undisturbed gas discs are preferentially found in intermediate mass haloes. The mass bins ($\mathrm{M1}$, $\mathrm{M2}$, $\mathrm{M3}$) comprise (47, 50, 3) percent of our fiducial sample, but correspond to (0.14, 75, 64) percent of the all subhaloes in the simulation in the corresponding mass bin. The pair counts of galaxies hosted by high-mass subhaloes declines sharply with decreasing separation distance owing to the low space density of massive haloes. 

\subsubsection{Auto-correlation}
\label{sec:mass_dep_auto}

Fig.~\ref{fig:mass_dep_auto} shows the orientation-direction alignment of the star-forming gas for subhalo pairs \textit{of similar mass}, with the lower panels showing the corresponding pair counts. The dashed black curves correspond to the $\mathcal{A}$ and $\mathcal{B}$ samples without mass binning (i.e. the blue curves from Fig.~\ref{fig:matter_dep}). Note that the dynamic range of the y-axis differs in the left and right panels. Sub-sampling the fiducial sample to obtain similarly massive pairs restricts the range of separations over which the intrinsic alignments can be examined, and yields somewhat noisy results. At absolute separations of $r\lesssim 1\Mpc$ the uncertainties are sufficiently large that the measured orientations for the M1 and M2 auto-correlations are consistent with a random distribution. All three bins are well sampled for $r\gtrsim 3\Mpc$, and on these scales it is clear that at fixed separation galaxies hosted by more massive subhaloes exhibit a more pronounced radial alignment. A similar trend for the mass dependence of the stellar component has been reported widely elsewhere \citep[e.g.][]{chisari15,tenneti15,vel15b}. As noted by \citet{vel15b} for the stars, normalising the pair separation by $r_\mathrm{DM}$ accounts for some, but not all, of the difference in star-forming gas alignment between mass bins. As is clear from the right panel of Fig.~\ref{fig:mass_dep_auto}, we similarly find that using this normalisation highlights that the radial alignment of M3 pairs becomes small at separations that are large compared to the half-mass radius of the primary galaxy ($r/r_{\mathrm{DM}} \simeq 30$).

\subsubsection{Cross-correlation}
\label{sec:mass_dep_cross}

\begin{figure}
\centering
\hspace{-0.2cm}
     \includegraphics[width = 0.5\textwidth]{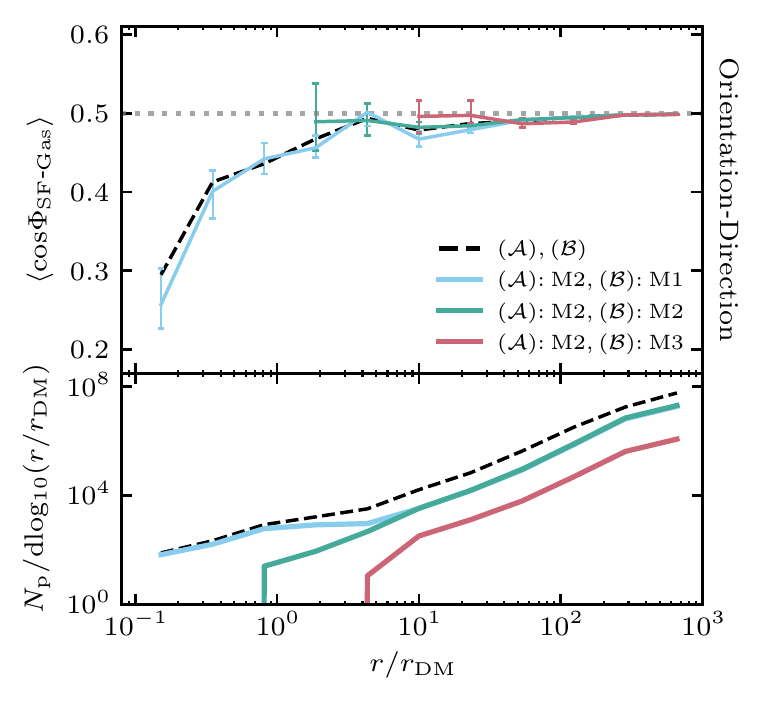}
\caption{The present-day orientation-direction alignment of the star-forming gas component of galaxy pairs for which the primary galaxy (sample $\mathcal{A}$) is drawn from the M2 bin and the secondary (sample $\mathcal{B}$) is drawn from the M1 (blue curves), M2 (green) or M3 (red) bins. The alignment is shown as a function of pair separation normalised by the DM half-mass radius of the primary galaxy's subhalo ($r_{\mathrm{DM}}$). The bottom row shows the corresponding pair counts. Dashed black curves correspond to the $\mathcal{A}$ and $\mathcal{B}$ samples without mass binning (i.e. the blue curves from Fig.~\ref{fig:matter_dep}). The dotted horizontal line corresponds to the expectation value for randomly-orientated 3-vectors (i.e. no intrinsic alignment). Error bars denote the bootstrap-estimated uncertainty on the measurements. Curves are drawn only for bins sampled by at least 10 galaxies. The strong orientation-direction alignments observed at short separations are dominated by pairings of $L_{\star}$ galaxies and their satellites.}
\label{fig:mass_dep_cross}
\end{figure}

As is clear from the lower panels of Fig.~\ref{fig:mass_dep_auto}, at small values of $r/r_{\mathrm{DM}}$ galaxy pairs of similar subhalo mass represent a small component of the total pair counts. We therefore next consider cross-correlations. Fig.~\ref{fig:mass_dep_cross} shows the orientation-direction alignment of galaxy pairs for the case in which the primary galaxy is drawn from the M2 bin, whilst the secondary galaxy is drawn from M1 (blue curves), M2 (green) or M3 (red). For brevity, we show only the case for which pair separations are normalised by $r_{\mathrm{DM}}$. By construction, the cross-correlation of equal mass (i.e. the auto-correlation, M2-M2) or more massive haloes (M2-M3) is ill-defined for short $r/r_{\mathrm{DM}}$ separations. The regimes that are sampled by all the considered cross-correlations exhibit only a very mild radial alignment, only marginally inconsistent with a random distribution for separations of $r/r_{\mathrm{DM}} \sim 10^1\mbox{-}10^2$. 

As is clear from the lower panel of Fig.~\ref{fig:mass_dep_cross}, at $r/r_\mathrm{DM} \lesssim 3$, the fiducial sample is dominated by M2-M1 pairs, and as seen in Fig.~\ref{fig:matter_dep} these pairs generally also share the same parent FoF halo.  At $r/r_{\mathrm{DM}} = (1, 0.2)$, M2-M1 pairs exhibit alignments of $\langle \cos\Phi \rangle = (0.445, 0.292)$. At these separations $\mathrm{M}2$-$\mathrm{M}1$ pairs represent $70$ percent and $85$ percent of all pairs. Unsurprisingly then, the M2-M1 orientation-direction alignment therefore closely mirrors that of the overall sample in this regime (illustrated by the cyan curve closely tracking the dashed black curve). Hence the EAGLE simulation indicates that the preferential radial alignment of the star-forming gas component of close galaxy pairs is driven largely by ${\sim}L^\star$ galaxies and their satellites.

\subsection{Intrinsic alignments as a function of redshift}
\label{sec:redshift_dep}

\begin{figure*}
\centering
\hspace{-0.2cm}
     \includegraphics[width = 1.\textwidth]{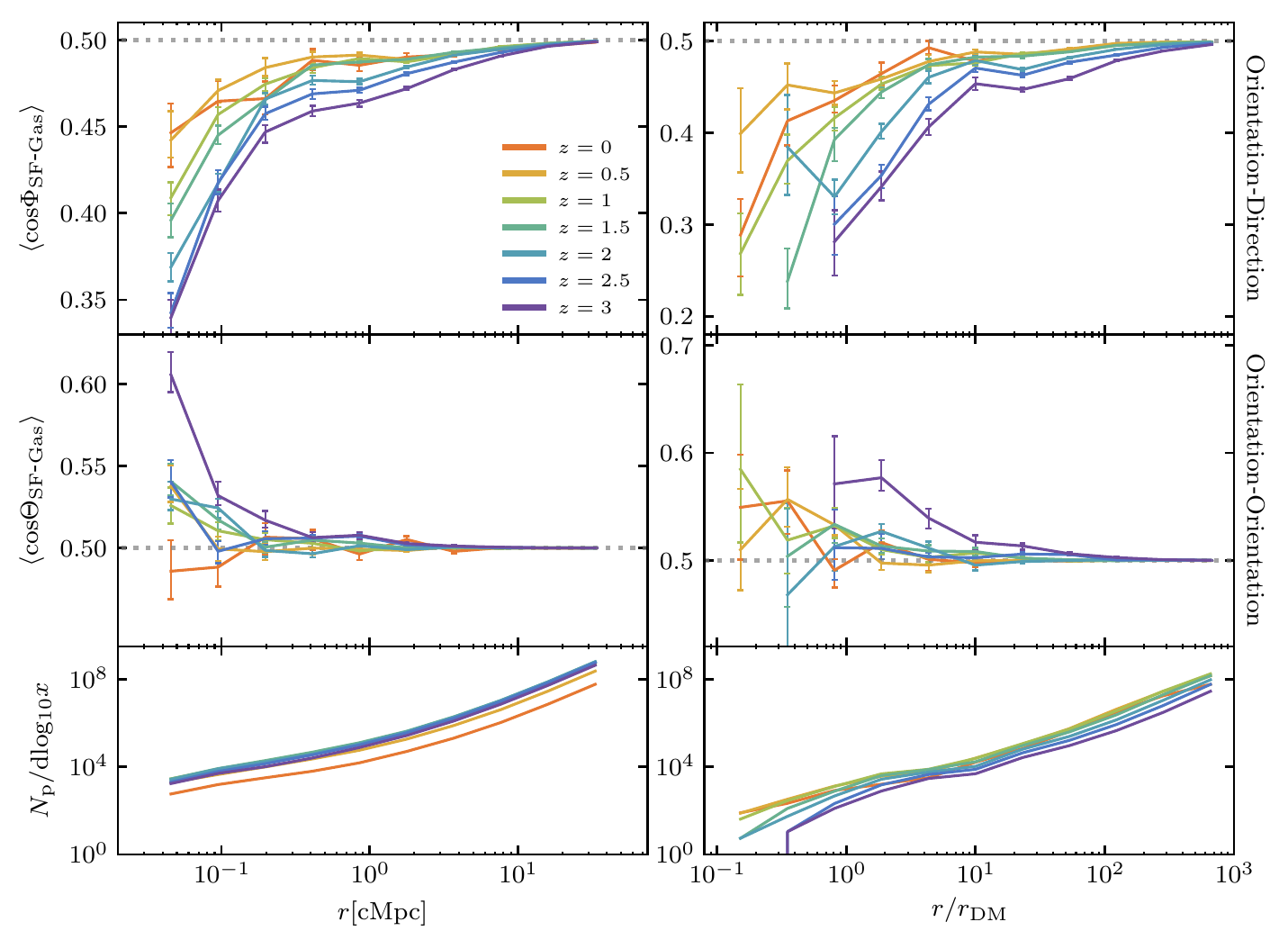}
\caption{The orientation-direction (top row) and orientation-orientation (middle row) intrinsic alignments of star-forming gas as a function of galaxy pair separation, at seven redshifts between $z=0$ and $z=3$, denoted by curve colour (see legend). Dotted horizontal lines correspond to the expectation value for randomly-orientated 3-vectors (i.e. no intrinsic alignment). The bottom row shows the corresponding pair counts. The left and right columns correspond, respectively, to the separation in absolute (comoving) space, and that normalised by the DM half-mass radius of the primary galaxy's subhalo ($r_{\mathrm{DM}}$). Error bars denote the bootstrap-estimated uncertainty on the measurements. Curves are drawn only for bins sampled by at least 10 galaxies. Orientation-direction alignment decreases with advancing cosmic time at fixed comoving separation. The orientation-orientation alignment is consistent with random except at high $z$.}
\label{fig:zdep_3d}
\end{figure*}

In \citetalias{hill21} we showed that the morphology of the star-forming gas bound to galaxies evolves with redshift, such that it exhibits increased flattening along the minor axis at later cosmic epochs. We therefore examine next whether there is a corresponding evolution of the intrinsic alignments as a function of redshift. This question is pertinent in the context of radio weak lensing surveys, which will obtain shape measurements for background galaxies at higher characteristic redshifts than their optical counterparts \citep{brown15,harrison,bonaldi,camera,ska_red}, and may therefore motivate a redshift-dependent intrinsic alignment mitigation strategy. 

We assess the 3-dimensional intrinsic alignments for star-forming gas at redshifts plausibly accessible to the SKA. As we are concerned only with star-forming gas here, we relax the requirement that subhaloes exhibit at least $100$ star particles. For context, we list key properties of the resulting sample at each redshift considered in Table~\ref{tab:sfg_props}. The typical number of particles with which we resolve the star-forming gas component is similar at all redshifts probed (${\sim}400$).

Fig.~\ref{fig:zdep_3d} shows the 3-dimensional intrinsic alignments, as a function of comoving galaxy pair separation, at seven redshifts spanning the range $z=0-3$. The orientation-direction alignment evolves markedly and in a largely monotonic fashion, such that at fixed separation the radial alignment is stronger at earlier times: at $r = 10~\mathrm{cMpc}$, $\langle \cos\Phi \rangle (z = 0,\ 1.5,\ 3) = (0.495,\ 0.496,\ 0.492)$; at $r = 1~\mathrm{cMpc}$, $\langle \cos\Phi \rangle (z = 0,\ 1.5,\ 3) = (0.486,\ 0.487,\ 0.465)$; and at $r = 0.1~\mathrm{cMpc}$, $\langle \cos\Phi \rangle (z = 0,\ 1.5,\ 3) = (0.465,\ 0.446,\ 0.409)$. The orientation-orientation alignment is weaker than the orientation-direction alignment at fixed separation, at all redshifts, being broadly consistent with random for pairs separated by $r>0.3~\mathrm{cMpc}$ or $r/r_{\mathrm{DM}} \gtrsim 10$. At early epochs, closely separated pairs exhibit a preference for parallel alignment of their minor axes.

\begin{table}
\centering

\bgroup
\def\arraystretch{1.5}
\begin{tabular}{l|rrrrr}
\hline
Redshift    & $N_{\mathrm{sub}}$ & $\log_{10} M_{\mathrm{sub}}$ & $\log_{10} M_{\star}$ & $\alpha_{\mathrm{all}}$ &  $r_{\mathrm{DM}}$  \\ 
 & & [$\Msun$] & [$\Msun$] & & [$\pkpc$] 
\\
\hline
$z = 0.0$	&	6766	&	$11.5^{+0.5}_{-0.4}$	&	$9.72^{+0.66}_{-0.53}$	&	$24.4^{+38.7}_{-15.6}$	&	$59.2^{+34.4}_{-30.0}$	\\
$z = 0.5$	&	13558	&	$11.2^{+0.5}_{-0.5}$	&	$9.40^{+0.75}_{-0.55}$	&	$30.6^{+36.8}_{-19.3}$	&	$25.3^{+15.0}_{-14.0}$	\\
$z = 1.0$	&	19784	&	$11.1^{+0.5}_{-0.5}$	&	$9.12^{+0.79}_{-0.54}$	&	$35.7^{+33.7}_{-21.7}$	&	$13.9^{+8.26}_{-7.71}$	\\
$z = 1.5$	&	22141	&	$11.0^{+0.5}_{-0.5}$	&	$8.91^{+0.78}_{-0.53}$	&	$36.7^{+32.2}_{-21.8}$	&	$9.17^{+5.30}_{-4.90}$	\\
$z = 2.0$	&	22245	&	$10.9^{+0.5}_{-0.5}$	&	$8.71^{+0.76}_{-0.51}$	&	$36.8^{+32.6}_{-21.2}$	&	$6.37^{+3.52}_{-3.21}$	\\
$z = 2.5$	&	21173	&	$10.8^{+0.5}_{-0.5}$	&	$8.56^{+0.75}_{-0.52}$	&	$36.8^{+32.1}_{-20.8}$	&	$4.82^{+2.55}_{-2.19}$	\\
$z = 3.0$	&	18421	&	$10.8^{+0.5}_{-0.5}$	&	$8.43^{+0.72}_{-0.52}$	&	$36.9^{+31.3}_{-21.1}$	&	$3.68^{+1.82}_{-1.57}$	\\
\hline
\end{tabular}
\egroup
\caption{Key properties of the galaxy sample, recovered using the less restrictive criteria described in Section \ref{sec:redshift_dep}, as a function of redshift. The table presents median values and the intervals to the $16^{\mathrm{th}}$ and $84^{\mathrm{th}}$ percentiles. Columns are as follows: the snapshot redshift, the sample size ($N_{\rm sub}$), the typical subhalo mass ($M_{\mathrm{sub}}$); the stellar mass ($M_{\mathrm{\star}}$); the misalignment angle in degrees between the star-forming gas and the (entire) dark matter subhalo ($\alpha_{\mathrm{all}}$); and the half-mass radius of the dark matter ($r_{\mathrm{DM}}$).}
\label{tab:sfg_props}
\end{table}

We generally recover greater alignment amplitudes for close pairs when normalising their separations by $r_{\mathrm{DM}}$, highlighting the important role of one-halo pairs in determining the overall alignment. At $r/r_{\mathrm{DM}} = 100$, $\langle \cos\Phi \rangle (z = 0,\ 1.5,\ 3) = (0.493, 0.493, 0.472)$; at $r/r_{\mathrm{DM}} = 10$, $\langle \cos\Phi \rangle (z = 0,\ 1.5,\ 3) = (0.478,\ 0.482,\ 0.453)$; and at $r/r_{\mathrm{DM}} = 1$, $\langle \cos\Phi \rangle (z = 0,\ 1.5,\ 3) = (0.440,\ 0.402,\ 0.283)$. We note that the horizontal shift of the $r_{\rm DM}$-normalised curves is driven in part by the growth of subhaloes. 

Since strong radial alignments at short separations are dominated by one-halo central-satellite pairs, we interpret the strong evolution in the orientation-direction alignment primarily as a horizontal shift of the curves, driven by the decreasing characteristic separation of galaxy pairs, both in terms of absolute comoving distance and with respect to the (growing) half-mass radius of the primarily galaxy's subhalo half-mass radius (see Table~\ref{tab:sfg_props}). The evolutionary behaviour of the two alignments is qualitatively similar to that exhibited by DM haloes \citep[e.g.][]{lee08, chen16}, but does not perfectly mimic the evolution of the DM component's alignments because, as shown in \citetalias{hill21}, star-forming gas is a relatively poor tracer of the DM structure. As shown in Table~\ref{tab:sfg_props}, the misalignment of the two components, characterised by $\alpha_{\rm all}$, is generally stronger at earlier epochs and likely leads to the intrinsic alignments of star-forming gas evolving less markedly than those of the DM over the same redshift range. We explore the impact of the alignment of the baryons and the DM of subhaloes on intrinsic alignments in greater detail in the next sub-section.

\subsection{Impact of internal galaxy-halo alignment on intrinsic alignments}
\label{sec:internal_alignment}

\begin{figure*}
\centering
\hspace{-0.2cm}
\includegraphics[width = 1.\textwidth]{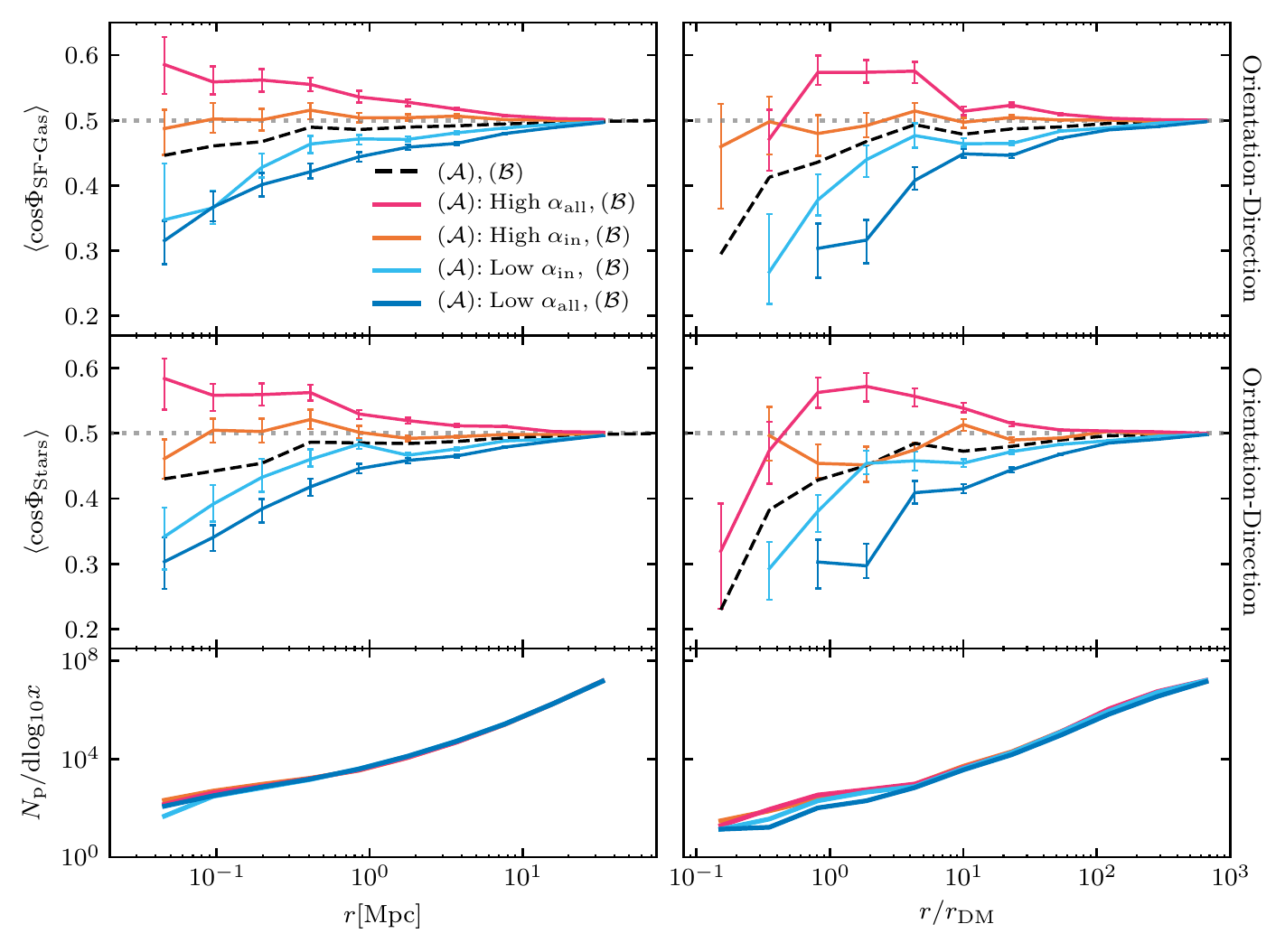}
\caption{The present-day orientation-direction intrinsic alignments of the baryonic components of galaxies (top row: star-forming gas, middle row: stars) as a function of galaxy pair separation. The bottom row shows the pair counts corresponding to the top row. Dashed black curves correspond to the fiducial galaxy sample, whilst coloured curves correspond to sub-samples with misalignment angles (defined by equation~\ref{eq:alpha}) between the relevant baryonic component and the subhalo DM that are either below the $25^{\mathrm{th}}$ percentile value (`Low $\alpha$', blue curves) or above the $75^{\mathrm{th}}$ percentile value (`High $\alpha$', red/orange shades). The misalignment angles are measured with respect to both the inner subhalo DM ($\alpha_{\rm in}$) and that of the entire subhalo ($\alpha_{\rm all}$). The left and right columns correspond, respectively, to the separation in absolute space, and that normalised by the DM half-mass radius of the primary galaxy's subhalo ($r_{\mathrm{DM}}$). Error bars denote the bootstrap-estimated uncertainty on the measurements. Curves are drawn only for bins sampled by at least 10 galaxies. Intrinsic alignments are strongly dependent on the internal alignment between baryons and the host dark matter halo.}
\label{fig:alignments}
\end{figure*}

The markedly different intrinsic alignments exhibited by the star-forming gas, stars and DM shown in Fig.~\ref{fig:matter_dep} imply that the different matter components within subhaloes can be poorly aligned. In \citetalias{hill21}, we showed that the distribution of present-day misalignment angles connecting the star-forming gas with the subhalo DM (see equation~\ref{eq:alpha}) peaks at low values ($<10^\circ$, i.e. good alignment) but exhibits a long tail to severe misalignments. That we find the star-forming gas to exhibit weaker intrinsic alignments than the stars is likely therefore a consequence of the former being a poorer tracer of the overall matter distribution, which is dominated by the DM. Such misalignment is clearly of consequence for weak lensing experiments: early studies of the auto-correlation of the intrinsic (stellar) ellipticities of galaxies found a lower amplitude than expected from theoretical predictions based on the assumption of perfect galaxy-halo alignment \citep{Heymans04, Mandelbaum06, Heymans06}. \citet{Okumura09} explored the impact of luminous red galaxy-host halo misalignment on the intrinsic ellipticity auto-correlation using $N$-body simulations, and concluded that the assumption of perfect galaxy-halo alignment results in predicted auto-correlation amplitudes four times higher than observed. 

We therefore assess the sensitivity of the orientation-direction alignment of galaxy pairs to the internal alignment of the baryonic components of the \textit{primary} galaxy and its subhalo DM, by constructing the $\mathcal{A}$ sample from sub-samples of galaxies that exhibit particularly good and particular poor internal alignment, characterised by the misalignment angle. We consider the misalignment of the star-forming gas and the stars with respect to the DM, and define well- and poorly-aligned systems, respectively, as those with misalignment angles below the $25^{\mathrm{th}}$ and above the $75^{\mathrm{th}}$ percentile values. To assess the influence of subhalo DM structure, we consider misalignment angles measured with respect to both the inner subhalo ($\alpha_{\mathrm{in}}$) and the subhalo in its entirety ($\alpha_{\mathrm{all}}$). For context, the sample boundaries for the misalignment of star-forming gas and the DM are $12^{\circ}$ and $49^{\circ}$ for $\alpha_{\mathrm{all}}$ and $5^{\circ}$ and $26^{\circ}$ for $\alpha_{\mathrm{in}}$. Note that the $\mathcal{B}$ sample remains comprised of the entire fiducial sample. 

The resulting orientation-direction alignments at $z=0$, as a function of pair separation, are shown in Fig.~\ref{fig:alignments}. The top row shows the effect on the orientation-direction alignment when sub-sampling based on the misalignment of the primary galaxy's star-forming gas and DM, whilst the middle row sub-samples based on the misalignment of the primary galaxy's stars and DM. The bottom row shows the pair counts corresponding to the top panel: these deviate from a simple one-quarter scaling of the pair counts for the fiducial sample only at short separations ($r\lesssim 100\kpc$ or $r\lesssim 3 r_\mathrm{DM}$). 

Whether one considers the star-forming gas or the stars, the orientation-direction alignment of galaxy pairs is clearly sensitive to the misalignment of the baryons with respect to the DM. Well-aligned galaxies (`Low $\alpha$', blue and cyan curves) exhibit a systematically stronger radial orientation-direction alignment (lower values of $\langle \cos\Phi \rangle$) than the fiducial sample (dashed black curves) at all separations. Conversely, galaxies with strong internal misalignment (`High $\alpha$', red and orange curves) exhibit systematically larger values of $\langle \cos\Phi \rangle$ than the fiducial sample at all pair separations. When binned by absolute separation, the `High $\alpha_{\rm in}$' sub-samples (defined using the misalignment of the DM with either the star-forming gas or the stars) are consistent with no intrinsic alignment at all separations, whilst the `High $\alpha_{\rm all}$' sub-samples exhibit \textit{tangential} alignment ($\langle \cos\Phi \rangle > 0.5$). Recalling that the orientation-direction alignment exhibited by subhaloes is stronger when one considers the entire subhalo rather than only its inner structure (see Fig.~\ref{fig:matter_dep}), it is unsurprising that the well- and poorly-aligned galaxies exhibit greater differences in their intrinsic alignment when defined using $\alpha_{\rm all}$ (red and blue curves) rather than $\alpha_{\rm in}$ (orange and cyan curves). 

The appearance of tangential orientation-direction alignment, i.e. the preference for the disc plane to be orthogonal to the direction to a neighbour, in galaxies with large misalignment angles is likely due to the minor axis of the baryonic component of these galaxies being well aligned with a different principal axis of the subhalo DM, rather than exhibiting poor alignment with any of the subhalo axes \citepalias[see Fig. 7 of][]{hill21}. It is interesting that, when binning by $r/r_\mathrm{DM}$, the `High $\alpha_{\rm all}$' sub-sample reverts to random orientation, or even radial alignment (as is the case when classifying misalignment based on the stars), at the short separations dominated by central-satellite pairs. However, we caution that for such close pairs, the DM structure of either or both of the subhaloes may deviate from axisymmetry as a result of tidal forces, and/or may be ill-defined as a consequence of the inability of purely 3-dimensional halo finding algorithms to identify sub-structures against the high background density of a parent halo \citep{Muldrew2011}. In either case the inferred subhalo orientation(s), and the corresponding misalignment angle, is compromised. 

\section{Projected alignment measurements}
\label{sec:2d_results}

\begin{figure*}
\centering
\hspace{-0.2cm}
     \includegraphics[width = 0.98\textwidth]{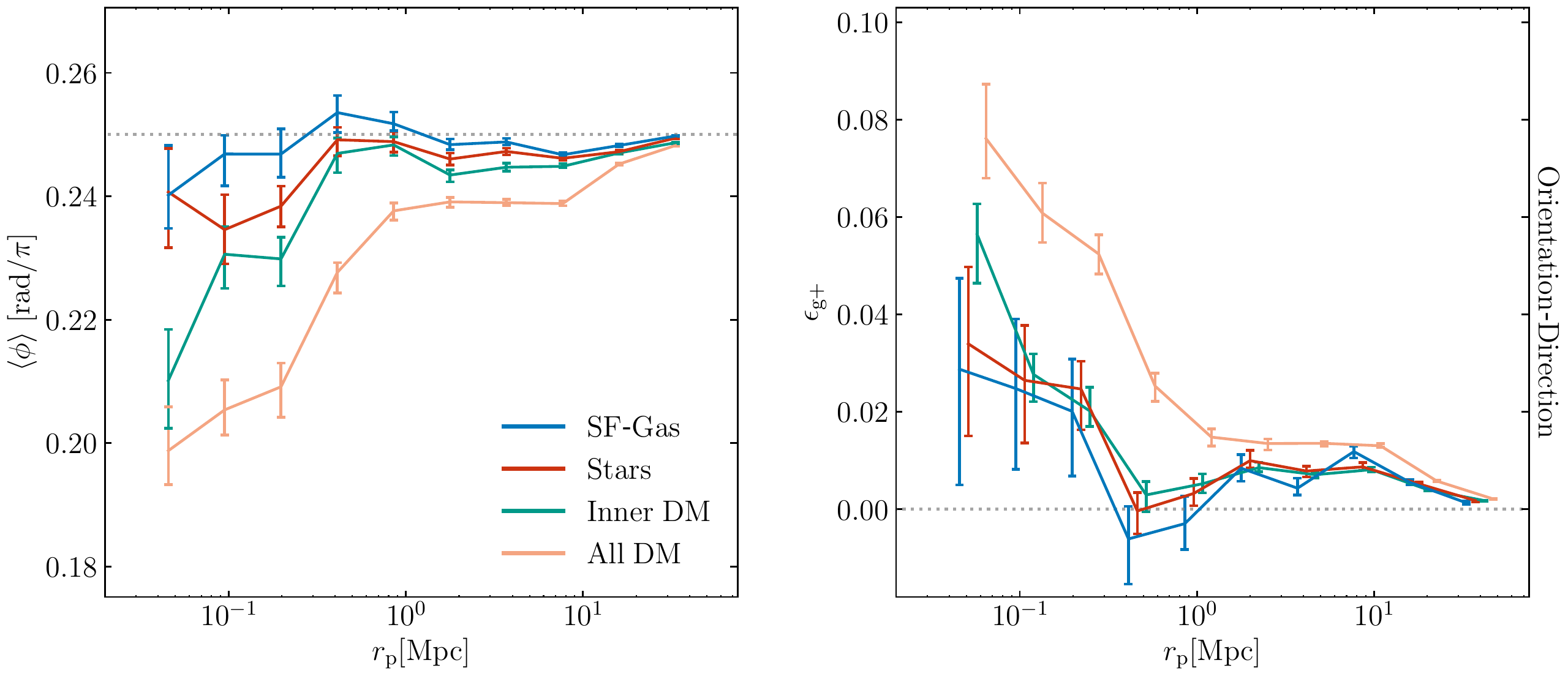}
\caption{The present-day 2-dimensional orientation-direction intrinsic alignments as a function of projected galaxy pair separation, for the star-forming gas (blue curves), stars (red) and DM (inner subhalo: green, entire subhalo: yellow) of our fiducial sample. The alignment is presented as the mean of the 2-dimensional alignment angle, $\phi$ (equation~\ref{eq:phi}), in the left panel, and as the mean intrinsic shear, $\epsilon_{\rm g,+}$ (equation~\ref{eq:eg_plus}), in the right column. Dotted horizontal lines correspond to the expectation values for randomly-orientated 2-vectors (i.e. no intrinsic alignment). Error bars denote the bootstrap-estimated uncertainty on the measurements. Curves are drawn only for bins sampled by at least 10 galaxies. For clarity, curves in the right panel are artificially offset along the x-axis by multiples of 0.05 dex. The projected orientation-direction alignment generally increases at decreased separation, and is weaker for the star-forming gas than the other matter components.}
\label{fig:matter_dep_2D_2p}
\end{figure*}

In this section we examine the projected orientation-direction and orientation-orientation alignments, mimicking the intrinsic alignments that act as sources of systematic uncertainty for observational weak lensing experiments. These quantities depend not only on the relative orientations of galaxies, but also on their projected morphology: more circular projected morphologies at fixed orientation result in lower ellipticity, $e_{+}$, and therefore reduced correlation function amplitudes. Authoritative prediction of the complex ellipticity therefore requires models with realistic galaxy morphologies. We showed in \citetalias[][]{hill21} (see their Fig. 11) that the projected star-forming gas morphologies of the galaxies comprising our sample are in good agreement those inferred by \citet{tunbridge} from Very Large Array (VLA) $L-$band observations of galaxies in the COSMOS field. For context, we remark that the `shear responsivity' values of our fiducial sample, defined as $\mathcal{R} = 1 - \langle e^{2}\rangle$ \citep{kaiser_squires, bernstein}, are $\mathcal{R_{\mathrm{SF\mbox{-}Gas}}}=0.59$ and $\mathcal{R_{\mathrm{stars}}}=0.83$, where the latter is comparable to the values obtained from analyses of SDSS data \citep[e.g.][]{sheldon,singh}.

\begin{figure}
\centering
\hspace{-0.2cm}
     \includegraphics[width = 0.47\textwidth]{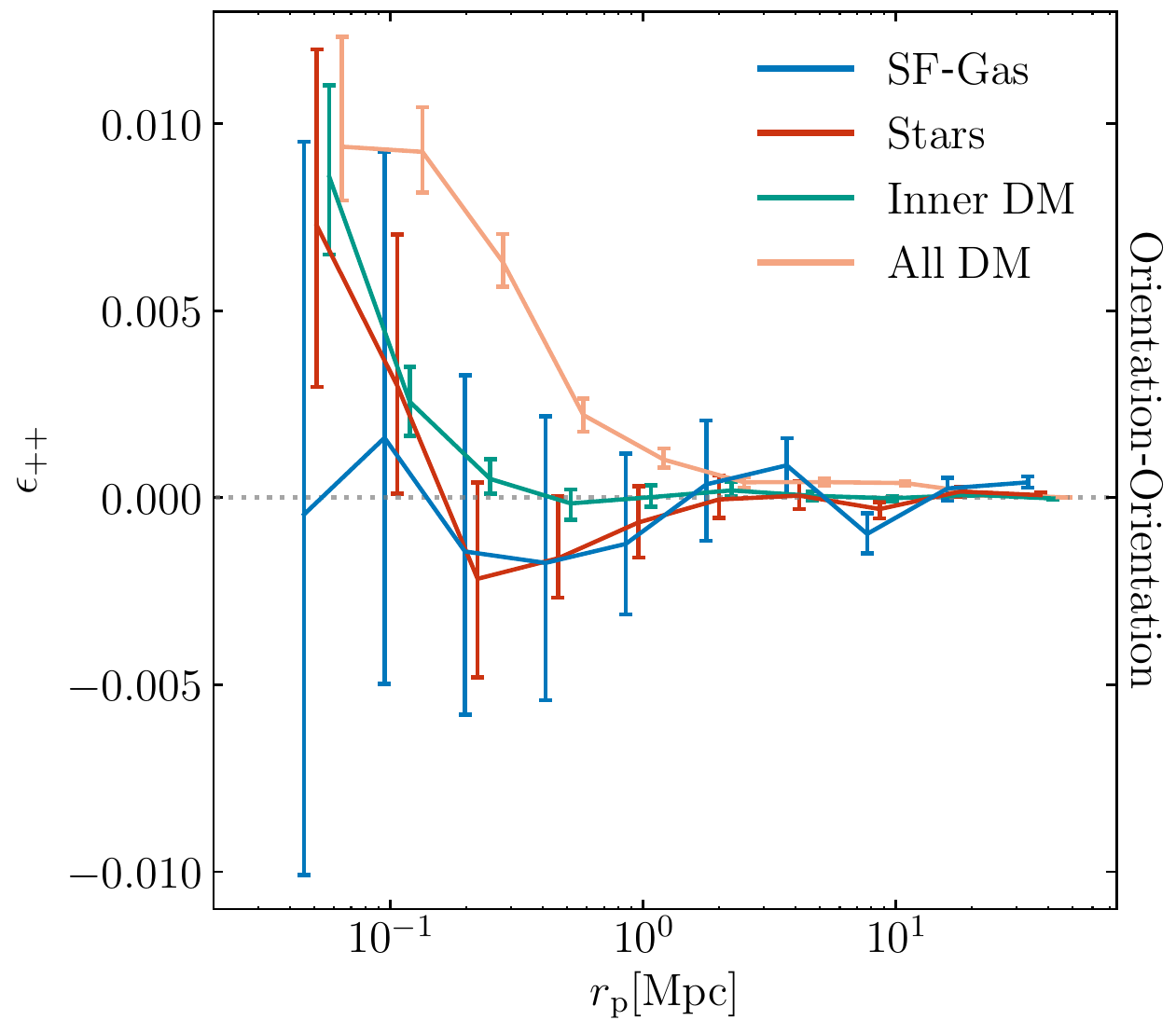}
\caption{The present-day 2-dimensional orientation-orientation alignment as a function of projected galaxy pair separation, for the star-forming gas (blue curves), stars (red) and DM (inner subhalo: green, entire subhalo: yellow) of our fiducial sample. The dotted horizontal line corresponds to the expectation values for randomly-orientated 2-vectors (i.e. no intrinsic alignment). Error bars denote the bootstrap-estimated uncertainty on the measurements. Curves are drawn only for bins sampled by at least 10 galaxies. For clarity, the curves are artificially offset along the separation axis by multiples of 0.05 dex. The orientation-orientation alignment of the star-forming gas is consistent with random at all separations.}
\label{fig:matter_dep_2D_1p}
\end{figure}

Fig.~\ref{fig:matter_dep_2D_2p} shows the projected orientation-direction alignment of the various matter components of galaxies. The left panel shows $\langle \phi \rangle$ (equation~\ref{eq:phi}), the mean angle subtended by the major axis of the primary galaxy's image and the direction vector to neighbouring galaxies (i.e. the 2-dimensional analogue of $\langle \Phi \rangle$). The expectation value for a random distribution of 2-vectors, $\pi/4~\mathrm{radians}$, is denoted by a dotted horizontal line. Values of $\langle \Phi \rangle$ below $\pi/4$ indicate a preference for radial alignment, i.e for the major axis of the image ellipse to point towards (projected) galaxy overdensities. We include this panel for ease of interpretation and to enable a more direct comparison with the 3-dimensional results presented in Fig.~\ref{fig:matter_dep}. The right panel shows the mean intrinsic shear for galaxy pairs ($\epsilon_{\mathrm{g+}}$, equation~\ref{eq:eg_plus}), for which the expectation value in the absence of intrinsic alignment is zero. Here, positive non-zero values indicate a preference for radial alignment.

In a similar fashion to the 3-dimensional case, all components exhibit an increasingly strong preferential radial alignment with decreasing separation. At fixed separation, the intrinsic alignment is strongest for the DM of the entire subhalo, followed in order by the DM of the inner subhalo, the stars and finally the star-forming gas. For the latter, $\langle \phi \rangle$ exhibits a significant deviation from $\pi/4$ only for pairs separated by $r\lesssim 100\kpc$, a markedly shorter scale than is the case for the stars ($r\lesssim 400\kpc$). The alignment we recover for the stars is broadly consistent with that inferred by \citet[][their Fig.~8]{vel15b}. Besides the difference in sample selection (since our fiducial sample is weighted towards star-forming galaxies), we note that \citet{vel15b} highlighted the particular sensitivity of $\epsilon_{\mathrm{g+}}$ (for the stars) to the choice of aperture used, which is in effect analogous to the application of a surface brightness limit.

The slightly greater statistical uncertainty on $\epsilon_{\mathrm{g+}}$ than $\langle \phi \rangle$ stems from the convolution of the projected morphology in the former. Although the 3-dimensional star-forming gas morphology exhibits a lower variance than the stars and DM, the converse is true for the projected morphology \citepalias[][see their Fig.~11 and the discussion therein]{hill21}. Despite these greater uncertainties and the generally poorer internal alignment of star-forming gas with the DM structure of subhaloes, the simulations indicate that a significant projected orientation-direction alignment of the star-forming gas component of galaxies is present for relatively close pairs. In contrast, the projected orientation-orientation alignment of the star-forming gas component ($\epsilon_{++}$, equation~\ref{eq:epp}), shown in Fig.~\ref{fig:matter_dep_2D_1p}, is consistent with random at all separations. This finding is perhaps unsurprising when one considers that the statistically significant orientation-orientation alignment of the entire subhalo DM at short separations, $\epsilon_{++,{\rm DM}}(r=0.1\Mpc) \simeq 0.01$, is a factor of several weaker than the corresponding orientation-direction alignment, $\epsilon_{g+,{\rm DM}}(r=0.1\Mpc) \simeq 0.75$. As such, it is unlikely that cosmic shear measurements in the radio continuum will be afflicted by a significant systematic error contributed by the II term. 

\section{Summary and discussion}\label{sec:discussion}

We have investigated the intrinsic alignments of the star-forming gas component of galaxies in the EAGLE suite of simulations \citep{schaye15, crain15, mcalpine16}. Our work is motivated by the need for authoritative theoretical predictions of the systematic uncertainties inherent to cosmic shear measurements conducted using radio continuum surveys which, with the forthcoming commissioning of the Square Kilometer Array \citep[SKA;][]{ska_red}, are poised to become competitive with, and complementary to, traditional optical weak lensing surveys. 

The realisation of these predictions requires state-of-the-art cosmological hydrodynamical simulations, which self-consistently follow the evolution of galaxies, their dark matter haloes and the cosmic large-scale structure, with spatial resolution on the order of $1\kpc$. They hence do not need to appeal to several of the most important assumptions and approximations inherent to the analytic and semi-analytic treatments of baryon physics used by galaxy alignment models, such as those relating the morphology and orientation of galaxies' baryonic components with respect to the dark matter of their host subhaloes. In the current state-of-the-art generation of hydrodynamical simulations of the galaxy population, fluid elements (i.e. gas particles or cells) with a non-zero star formation rate represent a good proxy for the interstellar gas that emits radio continuum radiation. EAGLE therefore represents an advantageous test-bed for this study, as many of the gaseous properties of its present-day galaxy population are broadly consistent with observations \citep[see e.g.][]{lagos15, bahe16, lagos16, crain17, dave20}.

We focus primarily on the present-day galaxy population, but also examine the evolution of intrinsic alignments over cosmic time. We examine the 3-dimensional orientation-direction, $\langle \cos\Phi \rangle$, and orientation-orientation, $\langle \cos\Theta \rangle$, alignments of galaxy pairs as a function of their separation, where the former is defined as the cosine of the angle between the minor axis of a galaxy and the direction vector to a neighbouring galaxy, and the latter is the cosine of the angle between the minor axes of neighbouring galaxies. To mimic the intrinsic alignments that potentially influence cosmic shear experiments, we also examine the corresponding alignments in 2-dimensions: the projected orientation-direction ($\epsilon_{\mathrm{g+}}$) and projected orientation-orientation ($\epsilon_{++}$) alignments.

Our results are summarised as follows:

\begin{enumerate}
    \item At fixed galaxy separation, the star-forming gas component of $z=0$ galaxies exhibits weaker intrinsic alignments in 3-dimensions than is the case for the stellar and dark matter (DM) components. Galaxy pairs, traced by any of these three matter components, exhibit an increasingly strong radial orientation-direction alignment at shorter separations. Radial orientation-direction alignment of the star-forming gas component persists even for pairs separated by 10s of Mpc, however the corresponding alignments for the stars and DM persist to greater separations still (Fig.~\ref{fig:matter_dep}). 
    
    \item In contrast, the star-forming gas component of galaxy pairs exhibits no significant orientation-orientation alignment at any separation, despite a significant preference for parallel alignment of the minor axes of DM subhaloes that persists out to separations of $r \sim 10\Mpc$ (Fig.~\ref{fig:matter_dep}). 

    \item We assess the mass dependence of the orientation-direction alignment by auto- and cross-correlating sub-samples of the fiducial sample defined by subhalo mass. The auto-correlation (Fig.~\ref{fig:mass_dep_auto}) reveals that, at absolute separations adequately sampled by galaxy pairs hosted by subhaloes with diverse masses, pairs of more massive subhaloes exhibit a more pronounced preference for radial alignment (at fixed separation), and this preference persists to greater separations. Normalising the pair separations by the half mass radius of the primary subhalo reduces, but does not eliminate, the mass dependence. Cross-correlating galaxy pairs when the primary galaxy is drawn from the intermediate subhalo mass bin reveals that the radial alignment of the fiducial sample is driven primarily by pairs comprising an ${\sim}L^\star$ galaxy and one of their satellites (Fig.~\ref{fig:mass_dep_cross}) .
    
    \item At fixed comoving separation, the orientation-direction alignment of galaxies' star-forming gas is greater at higher redshift, in a fashion qualitatively similar to that exhibited by the DM of subhaloes. We posit that this evolution is primarily a `horizontal' shift, i.e. the evolution of the characteristic separation of galaxy pairs dominates over the evolution of pairwise alignments. The orientation-orientation alignment is consistent with random for most redshifts and separations, however close pairs exhibit mildly preferential parallel alignment at early epochs (Fig.~\ref{fig:zdep_3d}).
    
    \item The orientation-direction alignment of star-forming gas is strongly influenced by the degree of misalignment between the star-forming gas and the DM structure of the galaxy's host subhalo. Galaxies whose star-forming gas is poorly aligned with the subhalo DM do not exhibit the radial orientation-direction alignment characteristic of the broader population. The most poorly-aligned galaxies (i.e. those with the largest internal misalignment angles) exhibit a preferential tangential alignment that increases with decreasing pair separation, likely as a consequence of the star-forming gas aligning more closely with either the intermediate or major, rather than the minor, axis of the subhalo's DM (Fig.~\ref{fig:alignments}). 

    \item The 2-dimensional orientation-direction alignments behave in a similar fashion to the 3-dimensional case, exhibiting increasingly preferential radial alignment at decreasing pair separations. The star-forming gas exhibits a weaker alignment at fixed separation that then stars and the DM, in turn (Fig~\ref{fig:matter_dep_2D_2p}). The projected orientation-orientation alignment of star-forming gas is consistent with random at all separations, despite their host DM subhalo exhibiting a preference for parallel alignment of the minor axes (Fig.~\ref{fig:matter_dep_2D_1p}). 
\end{enumerate}

In \citetalias{hill21} we showed that the characteristic morphology of the star-forming gas component of galaxies is a strong function of the mass of their host (sub)halo, and that the structure of the gas preferentially aligns with that of the subhalo's DM, albeit to a lesser degree than is the case for the stellar component. Here, we have shown that this internal alignment leads to the star-forming gas component of galaxy pairs exhibiting significant 3-dimensional orientation-direction alignment: the minor axis of a star-forming gas disc is preferentially perpendicular with respect to the direction vector connecting it with neighbouring galaxies, which can also be viewed as the plane of the disc pointing towards neighbouring galaxies. Viewed in projection, the 2-dimensional images of the discs, potentially visible as extended radio continuum emission, also exhibit an orientation-direction alignment that is strongest for close galaxy pairs.

However, we find that the intrinsic alignments of the star-forming gas component of galaxies are weaker (at fixed pair-separation) than the corresponding alignments of the galaxies' stars. This difference stems from the star-forming gas generally being a poorer tracer than the stars of the orientation and shape of the galaxies' DM structure. As such, we expect that the systematic uncertainty due to the intrinsic alignment of galaxies will have a milder influence on cosmic shear measurements conducted in the radio continuum regime than would be the case for an optical weak lensing survey over a similar redshift range. 

To our knowledge, the intrinsic alignments of star-forming gas have yet to be examined with traditional alignment models, in part owing to the complexity of realistically populating haloes with radio continuum sources. A promising avenue by which to estimate the intrinsic alignment uncertainty in radio continuum surveys may therefore be to adapt state-of-the-art simulations of the radio continuum sky \citep[see e.g.][]{wilman_08, bonaldi_19}. These simulations graft empirical or (semi-)analytic treatments of baryons onto $N$-body simulations of the large cosmic volumes needed to construct weak lensing survey lightcones, but cannot yet be used to model intrinsic alignments because, amongst other approximations, they assume that the radio continuum images of galaxies are oriented on the sky randomly. We caution against remedying this shortcoming by simply aligning the images with the projected structure of DM (sub)haloes, since we have shown that this leads to an overestimate of the star-forming gas intrinsic alignments. However, by relating the morphology and orientation of star-forming gas distributions to the properties of their host subhaloes, with careful reference to the corresponding relationships that emerge in state-of-the-art hydrodynamical simulations of the galaxy population, we envisage that it will be possible to use radio continuum sky simulations to predict the impact of intrinsic alignments on specific survey geometries. To this end, we note that analytic fits to the distribution functions star-forming gas misalignment angles, as a function of subhalo mass, are provided in \citetalias{hill21}.

\section*{Acknowledgements}
ADH is supported by an STFC doctoral studentship within the Liverpool Big Data Science Centre for Doctoral Training, hosted by Liverpool John Moores University and the University of Liverpool [ST/P006752/1]. RAC is a Royal Society University Research Fellow. This project has received funding from the European Research Council (ERC) under the European Union's Horizon 2020 research and innovation programme (grant agreement No 769130). The study made use of high performance computing facilities at Liverpool John Moores University, partly funded by the Royal Society and LJMU's Faculty of Engineering and Technology, and the DiRAC Data Centric system at Durham University, operated by the Institute for Computational Cosmology on behalf of the STFC DiRAC HPC Facility (www.dirac.ac.uk). This equipment was funded by BIS National E-infrastructure capital grant ST/K00042X/1, STFC capital grants ST/H008519/1 and ST/K00087X/1, STFC DiRAC Operations grant ST/K003267/1 and Durham University. DiRAC is part of the National E-Infrastructure.

\section*{Data Availability}

Particle data, and derived data products from the simulations have been released to the community as detailed by \citet{mcalpine16}. Further derived data used in this article will be shared on reasonable request to the corresponding author.


\bibliographystyle{mnras} 
\bibliography{lib1} 


\begin{appendix}

\section{Numerical convergence}
\label{sec:app_convergence}

\begin{figure}
\centering
\hspace{-0.2cm}
     \includegraphics[width = 0.46\textwidth]{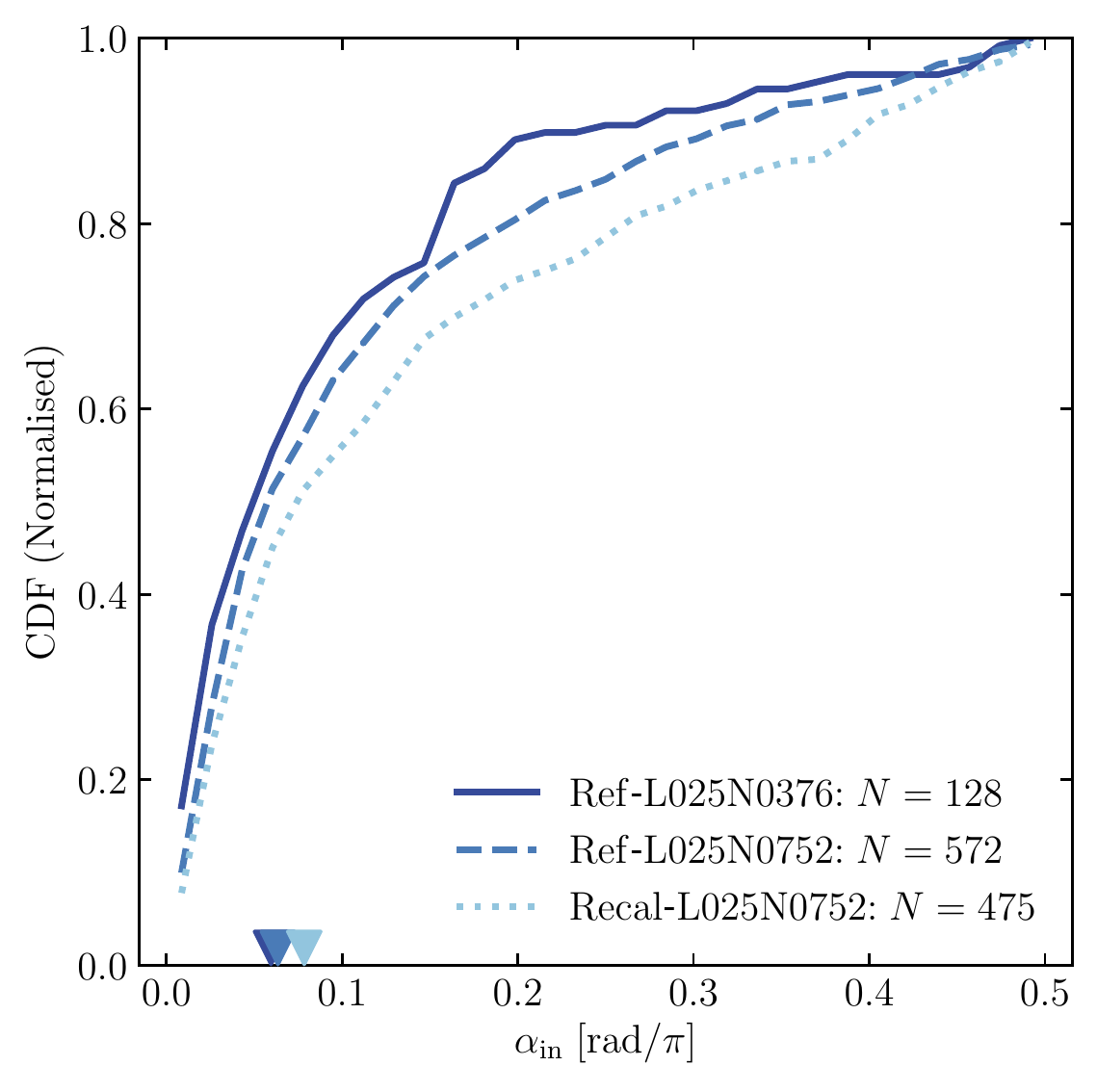}
\caption{Cumulative probability distribution functions of the present-day misalignment angle, $\alpha_{\mathrm{in}}$, between the minor axes of the star-forming gas of galaxies and the inner DM structure of their host subhaloes. These are drawn from the Ref-L025N0376 (solid dark blue curve), Ref-L025N0752 (dashed medium blue), and Recal-L025N0752 (dotted light blue) simulations. The number of galaxies satisfying the fiducial selection criteria is quoted in the legend. Down arrows denote the median values of each distribution. The similarity of the medians of each distribution compared to their interquartile ranges indicates that the misalignment angles are well converged in both the strong and weak senses.}
\label{fig:str_wk_conv}
\end{figure}

To assess the sensitivity of our findings to the numerical resolution of the Ref-L100N1504 simulation we examine three simulations from the EAGLE suite of a smaller $L=25\cMpc$ cosmological volume, also introduced by \citet{schaye15}. This enables direct comparison of the Reference model at EAGLE's fiducial resolution, Ref-L025N0376, with two higher-resolution simulations using particle masses a factor of eight lower. The first of these, Ref-L025N0752, again adopts the Reference model, enabling a test of what \citet[][see their Section 2]{schaye15} terms `strong convergence' (i.e. for a fixed model with changing resolution). The second, Recal-L025N0752, adopts a model recalibrated to achieve a better match to the calibration diagnostics at higher resolution, enabling a `weak convergence' test.

The number of galaxies satisfying the fiducial selection criteria (Section~\ref{sec:sampling}) in an $L=25\cMpc$ volume is too small to yield instructive measurements of orientation-direction alignment as a function of separation, so we focus here on the internal misalignment angle, $\alpha_{\mathrm{in}}$, subtended by the minor axes of the star-forming gas and the inner DM. As shown in Section \ref{sec:internal_alignment}, it is primarily this quantity that drives the difference in the intrinsic alignment of star-forming gas with respect to that of the subhalo DM. Fig. \ref{fig:str_wk_conv} shows the cumulative probability distribution function of $\alpha_{\mathrm{in}}$ for the subhaloes satisfying the fiducial selection criteria in the Ref-L025N0376 (solid curve), Ref-L025N0752 (dashed) and Recal-L025N0752 (dotted) simulations. Down arrows denote the median values of each distribution, which are (0.19, 0.2, 0.25)~radians, respectively. The differences between these median values are much smaller than the interquartile range of $\alpha_{\mathrm{in}}$ from any of the three simulations, e.g. for Ref-L025N0376 this range is 0.36. A similar trend is seen if one instead considers the misalignment angle between the minor axis of the star-forming gas, and that of the entire DM halo, $\alpha_{\mathrm{all}}$. The internal alignment angle $\alpha$ is therefore well-converged in both the strong and weak senses at the resolution of Ref-L100N1504, from which we infer that the star-forming gas intrinsic alignments are similarly well-converged.

\section{Influence of the subgrid ISM treatment}
\label{sec:app_subgrid}

\begin{figure}
\centering
\hspace{-0.2cm}
     \includegraphics[width = 0.46\textwidth]{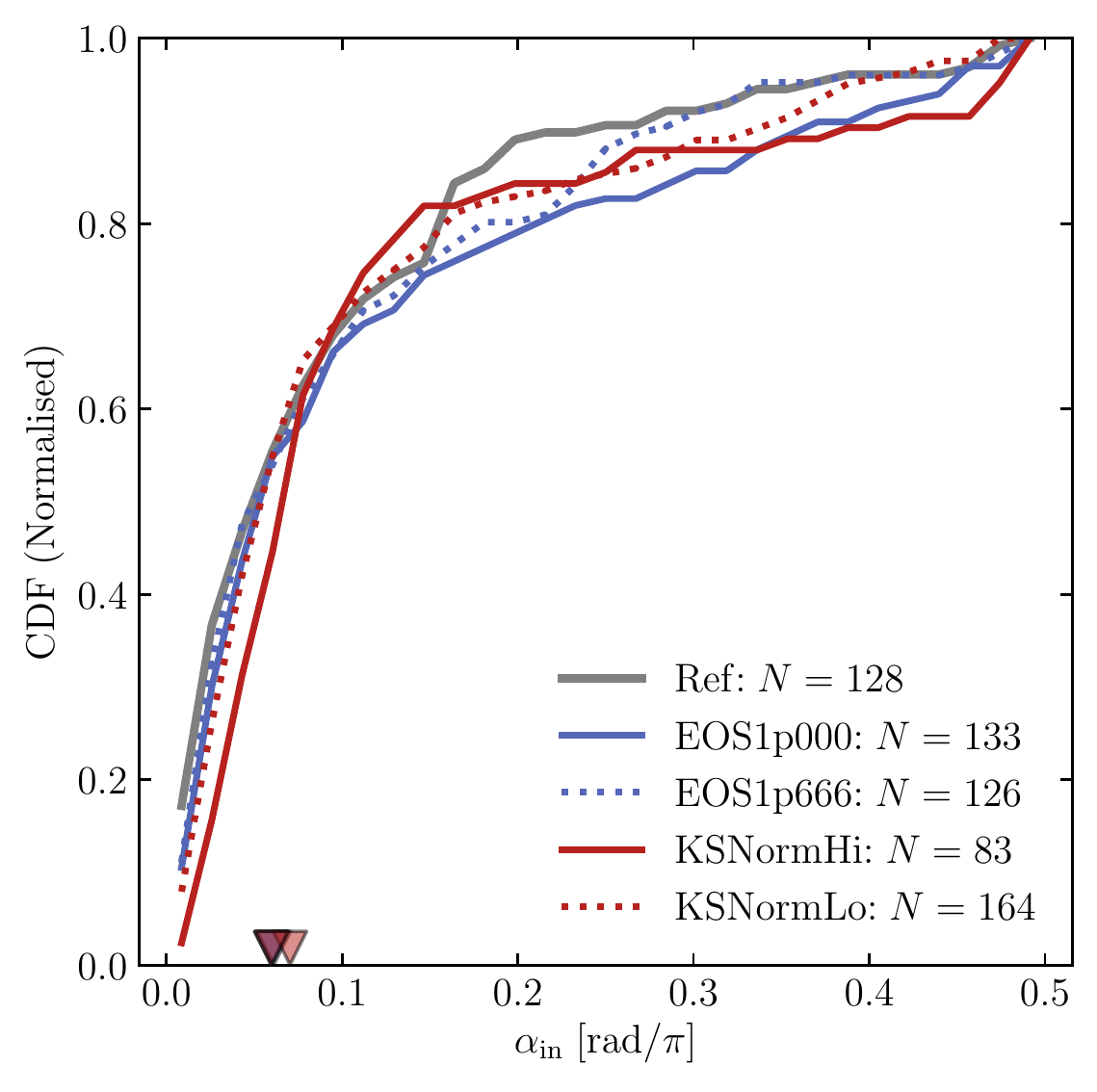}
\caption{Cumulative probability distribution functions of the present-day misalignment angle, $\alpha_{\mathrm{in}}$, between the minor axes of the star-forming gas of galaxies and the inner DM structure of their host subhaloes. These are drawn from the Ref-L025N0376 simulations (grey curve), and two pairs of simulations that incorporate variations of the reference model, with different slopes of the ISM equation of state (EOS1p00, solid blue; and EOS1p666, dotted blue) or normalisations of the relationship between the gas pressure and the star formation rate that differ from the reference value by $\pm 0.5$ dex (KSNormHi, solid red; KSNormLo, dotted red). The number of galaxies satisfying the fiducial selection criteria is quoted in the legend. Down arrows denote the median values of each distribution. The similarity of the medians of each distribution compared to their interquartile ranges indicates that the misalignment angles are robust to plausible changes to the subgrid physics of interstellar gas.}
\label{fig:eos_conv}
\end{figure}

We turn next to the sensitivity of alignments to the subgrid physics treatments in EAGLE that directly govern the properties of interstellar gas. We therefore compare results from Ref-L025N376 with those from two pairs of simulations of the same volume, and again consider the cumulative distribution function of  $\alpha_{\mathrm{in}}$. The first pair, introduced by \citet{crain15}, varies the slope of the equations of state (EoS) from the reference value of $\gamma_{\mathrm{eos}}=4/3$ to $\gamma_{\mathrm{eos}}=1$ (corresponding to an isothermal EoS) and $\gamma_{\mathrm{eos}}=5/3$ (an adiabatic EoS). \citet{schaye_DV} demonstrated that a stiffer EoS creates a smoother ISM with an increased scale height, and \citet{crain15} showed that a stiffer EoS also suppresses gas accretion onto central supermassive black holes in massive galaxies. The second pair, introduced by \citet{crain17}, varies the normalisation of the Kennicutt-Schmidt law \citep[the variable $A$ in equation 1 of][]{schaye15} from the reference value of $1.515\times10^{-4}\Msunyrsqkpc$ by $\pm 0.5$ dex. \citet{crain17} demonstrated that this parameter is inversely correlated with the mass of cold gas within galaxies, as it governs the gas mass needed to maintain an equilibrium between the rate of gas infall on one hand, and the rates of star formation and gas outflow due to ejective feedback on the other. 

Fig. \ref{fig:eos_conv} shows the cumulative probability distribution function of $\alpha_{\mathrm{in}}$ for Ref-L025N0376 (grey curve), the simulations with a differing EoS  ($\gamma_{\mathrm{eos}}=1$, solid blue; $\gamma_{\mathrm{eos}}=5/3$, dotted blue); and those with higher (solid red) and lower (dotted red) normalisations of the star formation law. As with the convergence test presented in Fig.~\ref{fig:str_wk_conv}, the distributions are not strongly affected by these significant changes to the subgrid physics: the median values of $\alpha_{\mathrm{in}}$ for each of the four variation simulations differ from that of Ref-L025N0376 by a maximum of $0.03$ radians, which is small compared to the interquartile range of the latter (0.36). Intrinsic alignments are therefore robust to plausible changes to the subgrid physics of interstellar gas.

\section{The influence of the adopted inertia tensor and aperture on inferred intrinsic alignments}
\label{sec:app_tensor}

\begin{figure*}
\centering
\hspace{-0.2cm}
     \includegraphics[width = 0.95\textwidth]{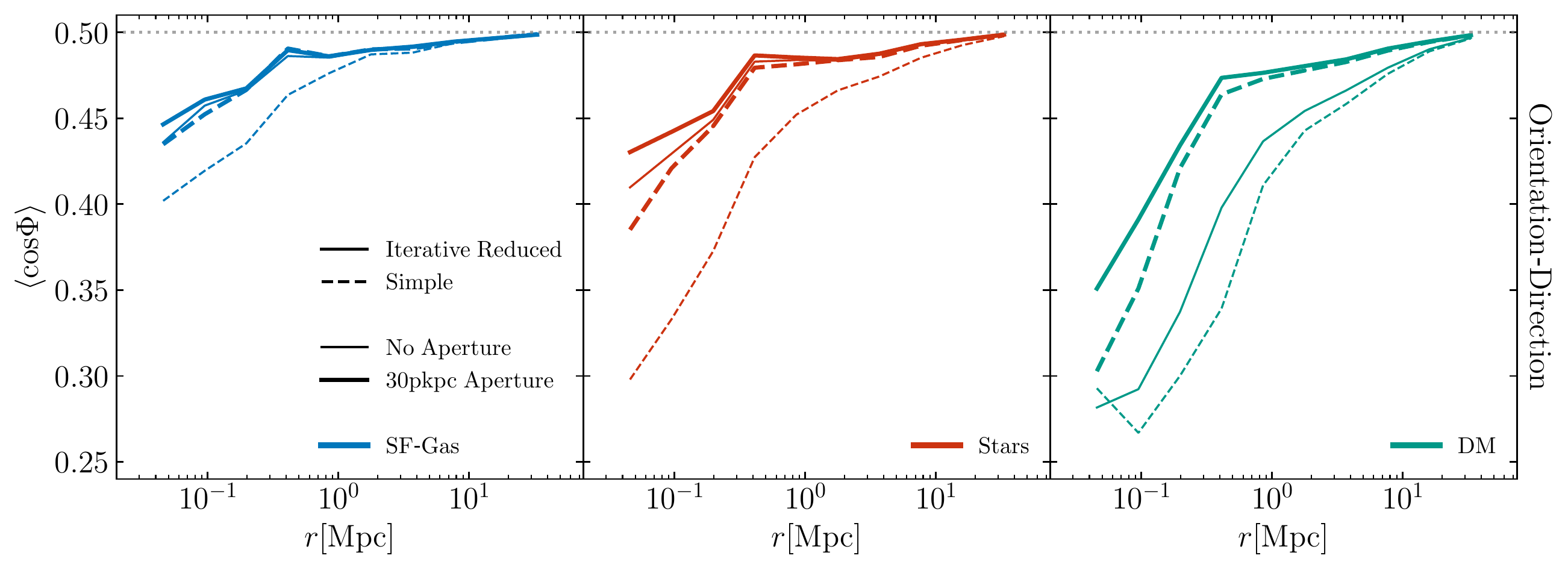}
\caption{The present-day 3-dimensional orientation-direction alignment as a function of galaxy pair separation. Displayed are the star-forming gas (left), stars (centre) and dark matter (right) components of galaxies in the Ref-L100N1504 simulation satisfying our fiducial selection criteria. Different curve styles and thicknesses correspond to different forms of inertia tensor (solid: iterative reduced, dashed: simple) and aperture ($30~\mathrm{pkpc}$: thick, no aperture: thin). The retrieved orientation-direction alignment of the star-forming gas is largely robust to the choice of shape-measurement algorithm, with the caveat that the simple inertia tensor with no aperture predicts larger alignments than the others at $r<1~\mathrm{Mpc}$.}
\label{fig:ia_test}
\end{figure*}

The adopted form of the inertia tensor can significantly influence the inferred morphology and orientation of the ellipsoid that best fits a particle distribution \citep[see e.g.][]{bett12}. Similarly, the choice of the aperture used to select the particles to be fitted has also been shown to markedly influence the inferred morphology and orientation of cosmic structures \citep[see e.g.][]{schneider,vel15b}. We therefore assess the sensitivity of the inferred 3-dimensional orientation-direction alignment to our use of an iterative form of the reduced inertia tensor, with an initially spherical aperture of radius $30\kpc$. Fig.~\ref{fig:ia_test} shows the alignment as a function of separation, recovered using both the reduced iterative inertia tensor (solid curves) and the simple inertia tensor (dashed curves), using both our standard aperture (thick curves) and no aperture (thin curves). From left to right, the three panels correspond to the star-forming gas, stars and dark matter, respectively.

This exercise reveals that that the qualitative trend inferred is the same in all cases, with the orientation-direction alignment of the star-forming gas being a decreasing function of pair separation. It is encouraging that, in a qualitative sense, the alignments of star-forming gas are much less sensitive to the choice of tensor and initial aperture than is the case for stars and the dark matter. We infer that this lower sensitivity stems from the more compact structure of the star-forming gas, which tends to be concentrated in subhalo centres. The deviation of the no aperture, simple tensor case from the other cases likely stems from the influence of isolated clouds of star-forming gas embedded in the circumgalactic medium of galaxies, some of which may be spurious \citep[see e.g.][]{schaller15b}.

\section{The influence of galaxy pair sampling on inferred intrinsic alignments}
\label{sec:app_sampling}

\begin{figure}
\centering
\hspace{-0.2cm}
     \includegraphics[width = 0.46\textwidth]{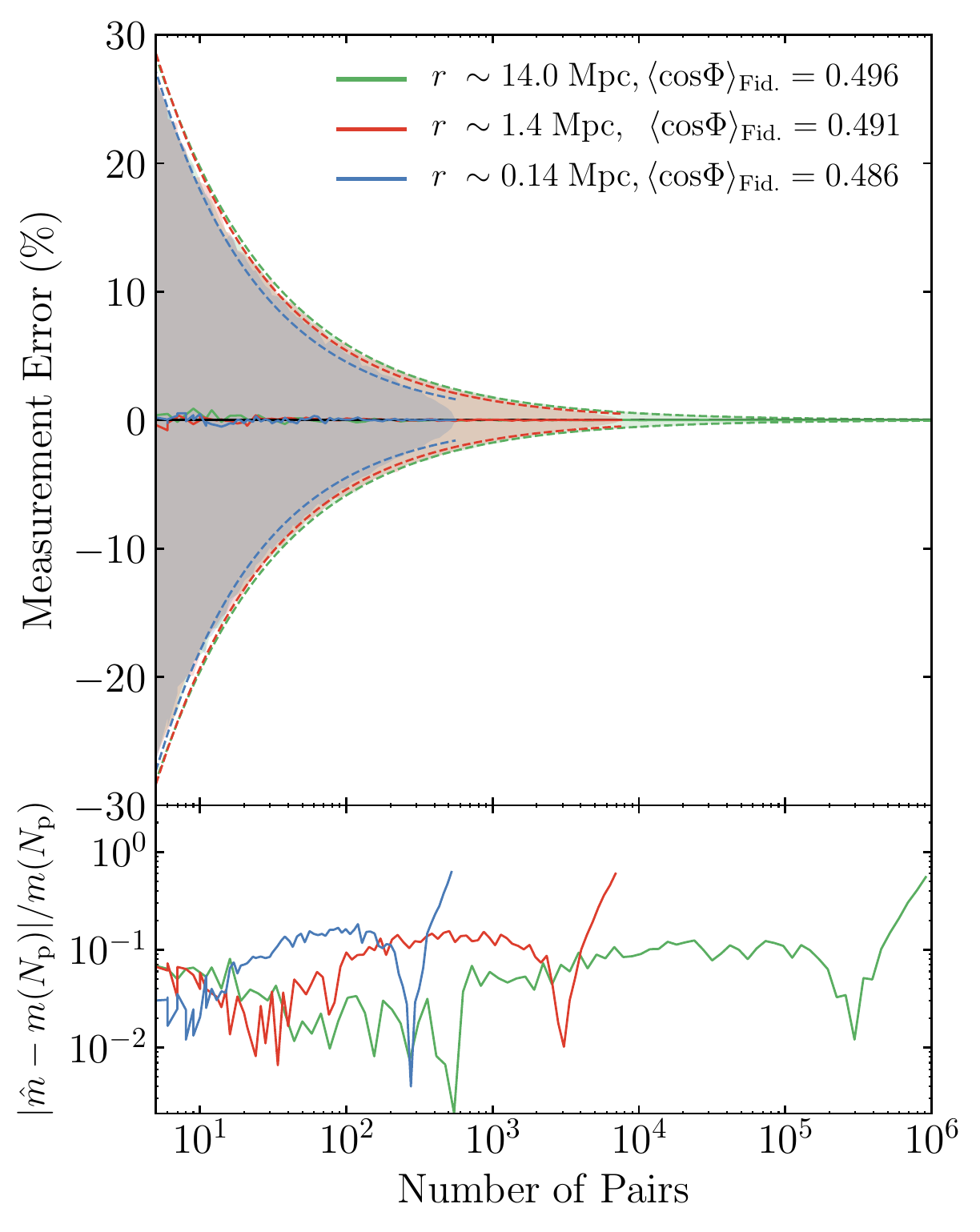}
\caption{The estimated fractional measurement error on the 3-dimensional orientation-direction alignment of star-forming gas, as a function of the number of galaxy pairs sampled. The estimates are obtained via 5000 measurements of $N_{\mathrm{p}}$ galaxies randomly drawn from the population of three well-sampled separation bins ($r {\sim}0.14, 1.4, 14\ \mathrm{Mpc}$, denoted by blue, red and green curves, respectively). Solid curves correspond to the median sampling error at fixed $N_{\mathrm{p}}$, and the shaded regions denote the interval bound by the $16^{\mathrm{th}}$ and $84^{\mathrm{th}}$ percentiles. The best-fit power laws are shown as dashed curves, whose residuals are shown in the lower panel.}
\label{fig:sampling_uncertainty}
\end{figure}

Measurement of the intrinsic alignments of simulated galaxies is unavoidably influenced by the finite sampling of galaxy pairs, particularly for short separations. We therefore obtain a basic estimate of the fractional uncertainty on inferred alignments as a function of the number of galaxy pairs used for the measurement, by recomputing the alignment of a well-sampled separation bin for sub-samples of the galaxy pairs. Fig.~\ref{fig:sampling_uncertainty} shows the fractional sampling error in $\langle\cos\Phi\rangle$ for the star-forming gas within three pair separation bins centred on $\sim$0.14, 1.4 and 14 Mpc. For each bin, we first compute a fiducial alignment measurement using all pairs, and then re-compute the measurement for 5000 randomly-drawn samples of $N_{\mathrm{p}}$ pairs. The dashed curves are functional fits to these bounds, calculated according to the power law
\begin{equation}
    m(N) = AN^{k}_{\mathrm{p}}\label{eq:fit},
\end{equation}
where $A$ and $k$ are free parameters, and $N_{\mathrm{p}}$ is the number of galaxy pairs. We calculate the best-fitting parameters with the Python package \textsc{scipy.optimize.curve\_fit}, and quote these in Table~\ref{tab:fit}.

The sampling error is roughly proportional to $1/\sqrt{N_{\mathrm{p}}}$, and is largely insensitive to the fiducial value of $\langle \cos\Phi \rangle$. Based on the best-fit associated with the $r=14.0~\mathrm{Mpc}$ separation bin, we find that the sampling error may be expected to fall below 1\% for $N_{\mathrm{p}} > 3000$, 5\% for $N_{\mathrm{p}}>140$ and 10\% for $N_{\mathrm{p}}>35$. We find similar results when repeating this test for the stars and the dark matter. 

\begin{table}
\centering

\bgroup
\def\arraystretch{1.5}
\begin{tabular}{l|rrr}
\hline
$r~[\mathrm{Mpc}]$          & 0.14        & 1.4           & 14 \\ \hline
$N_{\mathrm{tot}}$        & 558 & 7562 & 1047806 \\
$\langle \cos\Phi \rangle_{\mathrm{fid}}$        & 0.486 & 0.491 & 0.496 \\
$A_{84}$               & 70.9       & 69.7      & 65.7 \\
$k_{84}$               & -0.598     & -0.555    & -0.524 \\
$A_{16}$               & -72.0       & -69.0      & -65.2 \\
$k_{16}$               & -0.603     & -0.553    & -0.523 \\
\hline
\end{tabular}
\egroup
\caption{Best-fitting parameters of equation~\ref{eq:fit}, describing the the $16^{\rm th}$ and $84^{\rm th}$ percentiles of fractional measurement error estimated for the three separation bins ($r = 0.14, 1.4, 14\Mpc$) as a function of the number of galaxies sampled, $N_{\mathrm{p}}$. $\langle \cos\Phi \rangle_{\mathrm{fid}}$ is the fiducial measurement calculated using all $N_{\mathrm{tot}}$ galaxy pairs in each separation bin. The subscripts 16 and 84 on ($A,k$), the free parameters associated with equation~\ref{eq:fit}, denote the corresponding percentile being described.}
\label{tab:fit}
\end{table}

\end{appendix}
\label{lastpage}

\end{document}